\documentclass[10pt]{iopart}

\usepackage{graphicx}
\usepackage{color}
\usepackage{bm}
\usepackage{cite}
\usepackage{multirow}
\usepackage{hyperref}
\usepackage{braket}
\usepackage{bbm}

\definecolor{green}{rgb}{0,0.6,0.1}

\usepackage{iopams}  
\begin{document}

\title{Boltzmann machines and quantum many-body problems}

\author{Yusuke Nomura}

\address{Department of Applied Physics and Physico-Informatics, Keio University, 3-14-1 Hiyoshi, Kohoku-ku, Yokohama, 223-8522, Japan}
\ead{nomura@appi.keio.ac.jp}
\vspace{10pt}
\begin{indented}
\item[]January 2023
\end{indented}

\begin{abstract}
Analyzing quantum many-body problems and elucidating the entangled structure of quantum states is a significant challenge common to a wide range of fields.
Recently, a novel approach using machine learning was introduced to address this challenge.
The idea is to ``embed'' nontrivial quantum correlations (quantum entanglement) into artificial neural networks.
Through intensive developments, artificial neural network methods are becoming new powerful tools for analyzing quantum many-body problems. 
Among various artificial neural networks, this topical review focuses on Boltzmann machines and provides an overview of recent developments and applications.
\end{abstract}

%
%
%
%
\ioptwocol

\section{Introduction}
\label{Sec_introduction}

The motions of particles in a quantum many-body system are described by the following Schr\"odinger equation 
\begin{eqnarray}
\label{eq:many-body}
 {\mathcal H} | \psi \rangle = E | \psi \rangle 
\end{eqnarray}
where ${\mathcal H}$ is the Hamiltonian, an operator that represents the total energy of the system.
The specific form of the Hamiltonian depends on the type of particles constituting the quantum many-body system, how they interact with each other, and the form of the potential.

Despite the simplicity of the equation, its quantum and many-body nature gives rise to diverse quantum many-body phenomena that cannot be explained by the motion of a single particle. 
From a numerical perspective, the difficulty lies in the fact that the dimensions of the Hilbert space increase exponentially with the number of particles.
Therefore, obtaining exact solutions of the quantum many-body problems on a classical computer requires an exponentially large computational cost as the system size increases. 
To understand quantum many-body phenomena and design functional materials, it is necessary to develop powerful numerical methods that can obtain approximate but sufficiently accurate solutions with tractable computational resources.

To date, various numerical methods have been developed, including the quantum Monte Carlo (QMC) method and the variational wave function method.
The former applies the Monte Carlo method to estimate multidimensional integrals in quantum systems (for a review, see, e.g., Ref.~\cite{becca_sorella_2017}). 
If the sampling weights of the Monte Carlo method are always positive, it is possible to estimate the exact expectation values of 
physical quantities such as the energy and correlation functions within the margin of error.
However, in the case of challenging quantum many-body systems such as frustrated spin and fermion systems, a negative sign problem arises where some of the Monte Carlo sampling weights become unavoidably negative.
In the presence of a severe negative sign problem, an exponentially large computation time is required to accurately determine the true integral value obtained by summing the samples with positive and negative weights.
This limits the applicability of the QMC method. 
The latter approximates the eigenstate wave function of a quantum many-body system by a functional form described by, at most, a polynomial number of parameters.
In this case, physical quantities such as energy and correlation functions can be obtained without explicit negative sign problems, but their accuracy depends on the quality of the variational wave function.
Therefore, the problem is how to construct a powerful variational ansatz.

Historically, human insight has played a crucial role in the development of variational ansatzs. 
In fact, a variety of good ansatzs have been proposed, bringing innovation in our understanding of quantum many-body phenomena. 
Well-acknowledged wave functions include the Bardeen-Cooper-Schrieffer~\cite{Bardeen_1957} wave function for conventional superconductivity, the resonating valence bond wave function~\cite{Anderson_1973} for quantum spin liquid, and the Laughlin wave function~\cite{Laughlin_1983} for fractional quantum Hall effect.
Human-constructed wave functions have the advantage of being easy to interpret, but they have the disadvantage of being difficult to systematically improve the accuracy; it is highly nontrivial to identify the types of variational parameters that should be added for this purpose. 
Furthermore, for strongly correlated systems, as various quantum phases compete within a small energy scale, bias due to human insight may lead to qualitatively incorrect predictions.

Given this situation, it is interesting to see how machine learning techniques can contribute to this field. 
Using ``machine brains'' instead of human brains may open up new routes to building good variational ansatz.
The idea is to describe variational wave functions by means of artificial neural networks, exploiting the flexible representability of artificial neural networks. 
Notably, artificial neural networks have the property of universal approximation; 
in the limit of large networks, they can reproduce any wave function form with arbitrary accuracy~\cite{Clark_2018,Huang_2021}.
Thus, one can systematically improve the representative power of the variational ansatz by increasing the network size, which marks an advantage over insight-based ansatzs.

Such a new approach emerged in 2017 when Carleo and Troyer~\cite{Carleo_2017} introduced a variational wave function based on a restricted Boltzmann machine (RBM)~\cite{Smolensky_1986}. 
They showed that the RBM wave function can accurately simulate the ground state and real-time evolution of quantum spin systems without frustration~\cite{Carleo_2017}.  
Since then, extensions and improvements of the artificial neural network method have been continuously pursued all over the world.
Currently, its applications have been extended to simulations of
quantum spin systems with geometrical frustration~\cite{Cai_2018,Liang_2018,Choo_2019,Ferrari_2019,Westerhout_2020,Szabo_2020,Nomura_2021_JPCM,Nomura_2021_PRX,Astrakhantsev_2021,M_Li_2022,Rath_2022,Roth_2023,Reh_2023,Chen_arXiv}, 
itinerant boson systems~\cite{Saito_2017,Saito_2018}, 
fermion systems~\cite{Cai_2018,Nomura_2017,Luo_2019,Han_2019,Choo_2020,Pfau_2020,Hermann_2020,Stokes_2020,Yoshioka_2021,Inui_2021,Moreno_2022,Cassella_2023},
fermion-boson coupled systems~\cite{Nomura_2020}, 
topologically nontrivial quantum states~\cite{Deng_2017,Deng_2017_2,Glasser_2018,Clark_2018,Sirui_2019,Kaubruegger_2018,Huang_2021},
excited states~\cite{Choo_2018,Hendry_2019,Nomura_2020,Nomura_2021_JPCM,Vieijra_2020,Yoshioka_2021},
real-time evolution~\cite{Carleo_2017,Czischek_2018,Schmitt_2020},
open quantum systems~\cite{Nagy_2019,Hartmann_2019,Vincentini_2019,Yoshioka_2019},
and 
finite-temperature properties~\cite{Irikura_2020,Nomura_2021_PRL}.
In the early stages of research, most studies focused on benchmarks to check the accuracy of neural-network quantum states. 
More recently, calculations beyond benchmarks have begun to appear, and neural network methods are being applied to the analysis of unsolved problems, such as the physics of frustrated quantum spin systems~\cite{Nomura_2021_PRX,Astrakhantsev_2021} (note that the sign problem limits the applicability of the QMC method to frustrated spin systems).

Because the applications and architectures of neural-network quantum states are diverse, it is far beyond the scope of this article to cover all the developments.
Here, we focus on Boltzmann machines among various neural network architectures and review recent developments in quantum state representation using Boltzmann machines.
The remainder of this paper is organized as follows. 
In Sec.~\ref{Sec_BM}, we introduce Boltzmann machines as generative models. 
Then, in Sec.~\ref{Sec_BM_WF}, we discuss how to use Boltzmann machines to construct variational wave functions to be applied to quantum many-body problems. 
Secs.~\ref{Sec_RBM_zero}--\ref{Sec_DBM_finite} are devoted to more concrete applications. 
Sec.~\ref{Sec_RBM_zero} discusses the variational approach using RBMs for zero temperature calculations. 
Secs.~\ref{Sec_DBM_zero} and \ref{Sec_DBM_finite} introduce algorithms using deep Boltzmann machines (DBMs) for zero-temperature and finite-temperature simulations, respectively. 
Finally, we give a summary and outlook in Sec.~\ref{Sec_Summary}.

\section{Boltzmann machines} 
\label{Sec_BM}

Here, we introduce the RBM and DBM as generative models.
The generative model is a generic term for mathematical models that are intended to model the structure and probability distribution inherent in data.
In the next section (Sec.~\ref{Sec_BM_WF}), we will discuss how these Boltzmann machines can be applied to modeling ``quantum data,'' or quantum states.

\subsection{Restricted Boltzmann machines}

\begin{figure}[tbp]
\vspace{0cm}
\begin{center}
\includegraphics[width=0.50\textwidth]{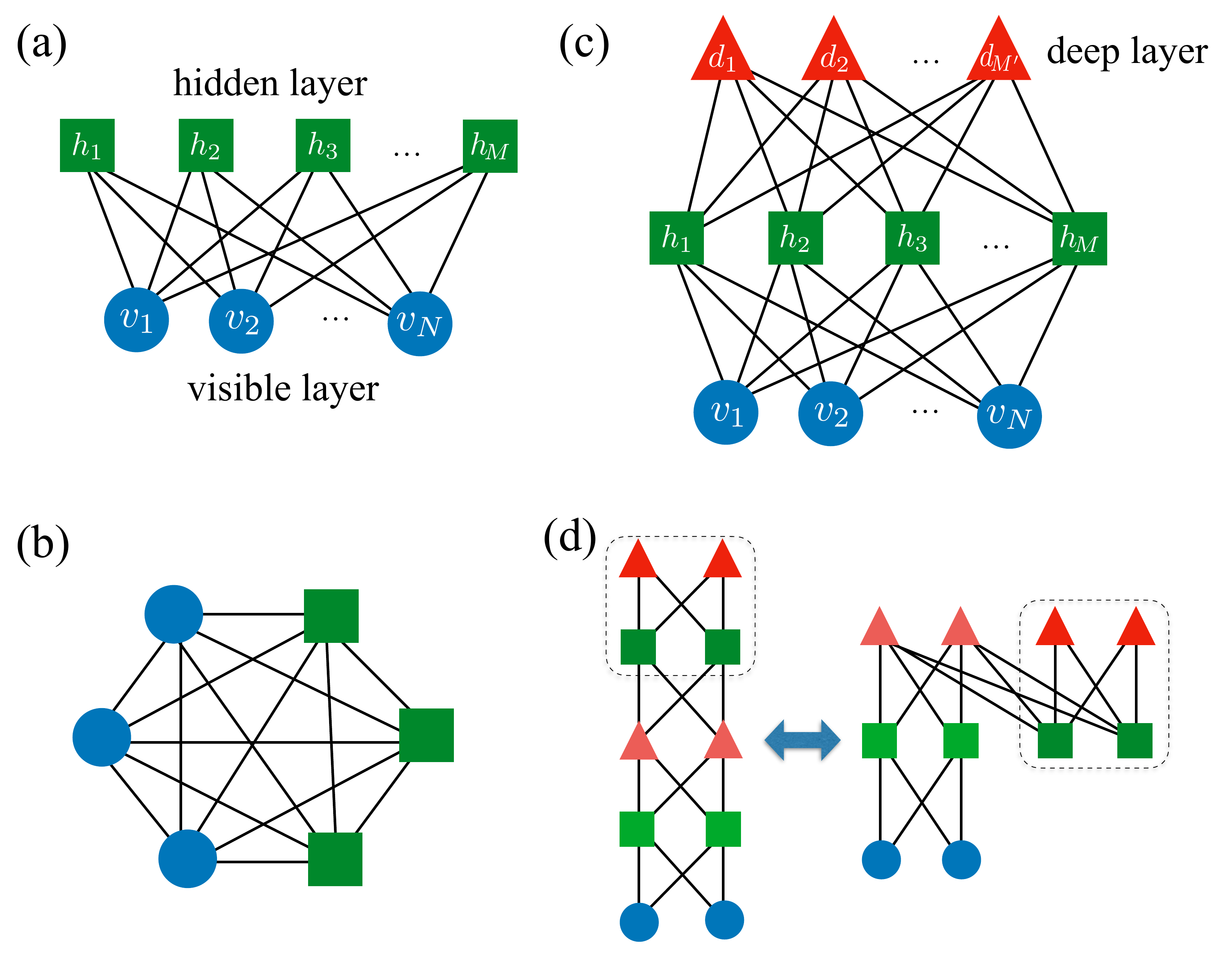}
\caption{
Structures of (a) restricted Boltzmann machine (RBM), (b) general Boltzmann machine, and (c) deep Boltzmann machine (DBM).
(d) Structures with more than two hidden layers can be attributed to structures with two hidden layers by changing the position of the hidden units.
}
\label{Fig_structure_BM}
\end{center}
\end{figure}

Fig.~\ref{Fig_structure_BM}(a) shows the structure of the RBM. 
The RBM is composed of visible units $\{v_i\}$ that make up the data space and extra degrees of freedom called hidden units $\{h_j \}$. 
The regions where each degree of freedom exists are called visible and hidden layers. 
Both $v_i$ and $h_j$ take two states; here we define the values of the two states as $v_i = \pm 1$, $h_j = \pm 1$
(another popular definition of the two states is $v_i, h_j=0$, 1).
When discussing applications to physics, it is convenient to identify them with Ising spins; therefore, we will refer to them as visible and hidden spins in this paper.

The RBM approximates probability distributions over visible spin configurations, as described below.
First, we define the artificial energy function $E(v, h)$ over the visible spin configuration $v=(v_1,\ldots,v_N)$ and the hidden spin configuration $h=(h_1,\ldots,h_M)$ as  
\begin{eqnarray}
\label{Eq.ene_function_RBM}
   E(v, h)  =   - \sum_{i} a_i v_i   - \sum_{i,j}  W_{ij} v_i h_j  - \sum_{j} b_j h_j . 
\end{eqnarray}
Here, $a_i$ and $b_j$ are the magnetic fields on the visible and hidden spins, respectively, and $W_{ij}$ is the classical interaction between the visible and hidden spins. 
In the machine learning field, the terms associated with $a_i$ and $b_j$ are called bias terms.
For the artificial energy $E(v, h)$, we can define the Boltzmann weight $p(v, h) = \frac{ e^{ - E(v, h)}}{Z}$, where $Z$ is the partition function $Z  = \sum_{ v, h } e^{-E(v, h)}$.
By tracing out the hidden-spin degrees of freedom, we obtain the marginal probability distribution $\tilde{p} (v)$: 
\begin{eqnarray}
\label{Eq_RBM_marginal}
\tilde{p} (v) = \sum_h p(v,h)  \quad  \quad   \Bigl ( \sum_{v}  \tilde{p} (v)  = 1 \Bigr  ).
\end{eqnarray}
Thus, the RBM takes the visible spin configuration $v$ as the input and gives the marginal probability $\tilde{p}(v)$ as the output.
This $\tilde{p}(v)$ is used to approximate the probability distributions over the visible spin configurations.

This Boltzmann machine is called ``restricted'' Boltzmann machine because of the restriction that there is no intralayer interaction in the visible and hidden layers [For comparison, we show the structure of a general Boltzmann machine in Fig.~\ref{Fig_structure_BM}(b)]. 
Because there is no interaction between hidden spins, one can analytically trace out hidden spins, and $\tilde{p}(v)$ in Eq.~(\ref{Eq_RBM_marginal}) can be efficiently computed as
\begin{eqnarray}
\label{Eq_analytic_marginal}
\tilde{p} (v) \propto  \exp \Bigl( \sum_{i} a_i v_i \Bigr) \times  \prod_j   2 \cosh \Bigl( b_j + \sum_{i} W_{ij}  v_i  \Bigr ).
\end{eqnarray}
As briefly mentioned in Sec.~\ref{Sec_introduction}, artificial neural networks have the property of universal approximation. 
In the case of the RBM, an arbitrary probability distribution can be reproduced with arbitrary precision by introducing an exponentially large number of hidden spins ($M\sim 2^N$) comparable to the number of visible spin configuration patterns $2^N$~\cite{Le_Roux_2008,Montufar_2011}.

\subsection{Deep Boltzmann machines}
\label{sec_DBM_explanation}

The DBM has an additional hidden layer compared to the RBM [Fig.~\ref{Fig_structure_BM}(c)]. 
The structure with two hidden layers is general, encompassing structures with an arbitrary number of hidden layers with only neighboring interlayer couplings [Fig.~\ref{Fig_structure_BM}(d)].
In this paper, the two hidden layers are referred to as the hidden layer and the deep layer for the sake of distinction. 
In the case of DBM, the artificial energy $E(v, h, d)$ with respect to the spin configurations $v=(v_1,\ldots,v_N)$, $h=(h_1,\ldots,h_{M})$, and $d=(d_1,\ldots,d_{M'})$ reads
\begin{eqnarray}
\label{Eq.ene_function_DBM}
   E(v, h, d)  =   &-&  \sum_{i} a_i v_i   - \sum_{j} b_j h_j -  \sum_{k} b'_k d_k    \nonumber \\ 
                     &-&    \sum_{i,j}  W_{ij} v_i h_j  - \sum_{j,k} W'_{jk} h_j d_k. 
\end{eqnarray}
In addition to the parameters in the RBM, the DBM has parameters for the magnetic field on the deep spins $b'_k$ and the interaction between the hidden and deep spins 
$W'_{jk}$. 
As with the RBM, using the partition function $Z  = \sum_{ v, h,d } e^{ -  E(v, h,d)  }$ and the Boltzmann weight $p(v,h,d)  = \frac{ e^{-  E(v, h,d) }  } { Z } $, the marginal probability distribution $\tilde{p} (v)$ is given by
\begin{eqnarray} 
 \tilde{p} (v) = \sum_{h,d}  p(v,h,d)   \quad  \quad   \Bigl ( \sum_{v}  \tilde{p} (v)  = 1 \Bigr  ). 
 \end{eqnarray}
Compared to the RBM, the DBM has the advantage of having deep degrees of freedom, which greatly improves the ability to represent the probability distribution, but since there is no efficient analytical formula for $\tilde{p}(v)$~\footnote{Either hidden or deep spin degrees of freedom can be traced out analytically, but tracing out spins of one layer will introduce effective interactions between the spins of the other layer. This makes it difficult to analytically trace out the spin degrees of freedom of the other layer.}, numerical methods such as the Monte Carlo method are required to obtain the $\tilde{p}(v)$ value.

\section{Wave function based on Boltzmann machines}
\label{Sec_BM_WF}

\subsection{Application of Boltzmann machines to quantum state representation}

Here, we introduce how Boltzmann machines can be used to represent quantum states. 
A quantum state $| \psi \rangle$ can be expanded using a certain basis $\{| x \rangle \}$ as 
\begin{eqnarray}
   | \psi \rangle = \sum_x  | x \rangle  \psi(x), 
\end{eqnarray}
where $\psi(x)$ is quantum many-body wave function. 
As seen in the previous section (Sec.~\ref{Sec_BM}), Boltzmann machines give a probability (positive real number) $\tilde{p} (v)$ for a visible spin configuration $v$.
On the other hand, the amplitude of quantum many-body wave function $\psi(x)$ can be negative or complex in general.

Therefore, we can apply Boltzmann machines to quantum state representations by satisfying the following conditions: 
\begin{enumerate}
\item One-to-one correspondence between the configuration of physical degrees of freedom $x$ and visible-spin configuration $v$. 
\item Extension to complex amplitude.
\item (For fermion systems) Incorporation of fermionic anti-commutation relations. 
\end{enumerate}

For condition (i), one needs to define the mapping depending on the nature of the physical degrees of freedom. 
For $S=1/2$ quantum spin systems, the most natural mapping between $x$ and $v$ is to identify the $z$ component of $i$-th quantum spin $\sigma_i^z$ with $v_i$. 

Condition (ii) can be satisfied, e.g., by extending the parameters of Boltzmann machines to complex variables or by decomposing the wave function as $\psi(x) = | \psi(x) | e^{i \phi (x)}$ and expressing $|\psi(x)|$ and $\phi(x)$ using independent Boltzmann machines with real parameters. 
For example, when approximating the quantum states of the one-dimensional Heisenberg model by the RBM,  
the former appears to perform better~\cite{Viteritti_2022}. 
We note however that, for other general Hamiltonians, it is unclear whether the former is always better.
Hereafter, we will focus on the former approach. 

For condition (iii), for example, one can take an approach to make the wave function itself anti-symmetric.
Alternatively, one can consider an approach that maps fermionic systems into spin systems using, e.g., Jordan-Wigner~\cite{Jordan_1928} or Bravyi-Kitaev~\cite{Bravyi_2002} transformations.  
This issue will be discussed in more detail in Sec.~\ref{sec_fermion}.

\subsection{Property of Boltzmann machine wave functions and comparison to tensor networks}

\begin{figure}[tbp]
\vspace{0cm}
\begin{center}
\includegraphics[width=0.48\textwidth]{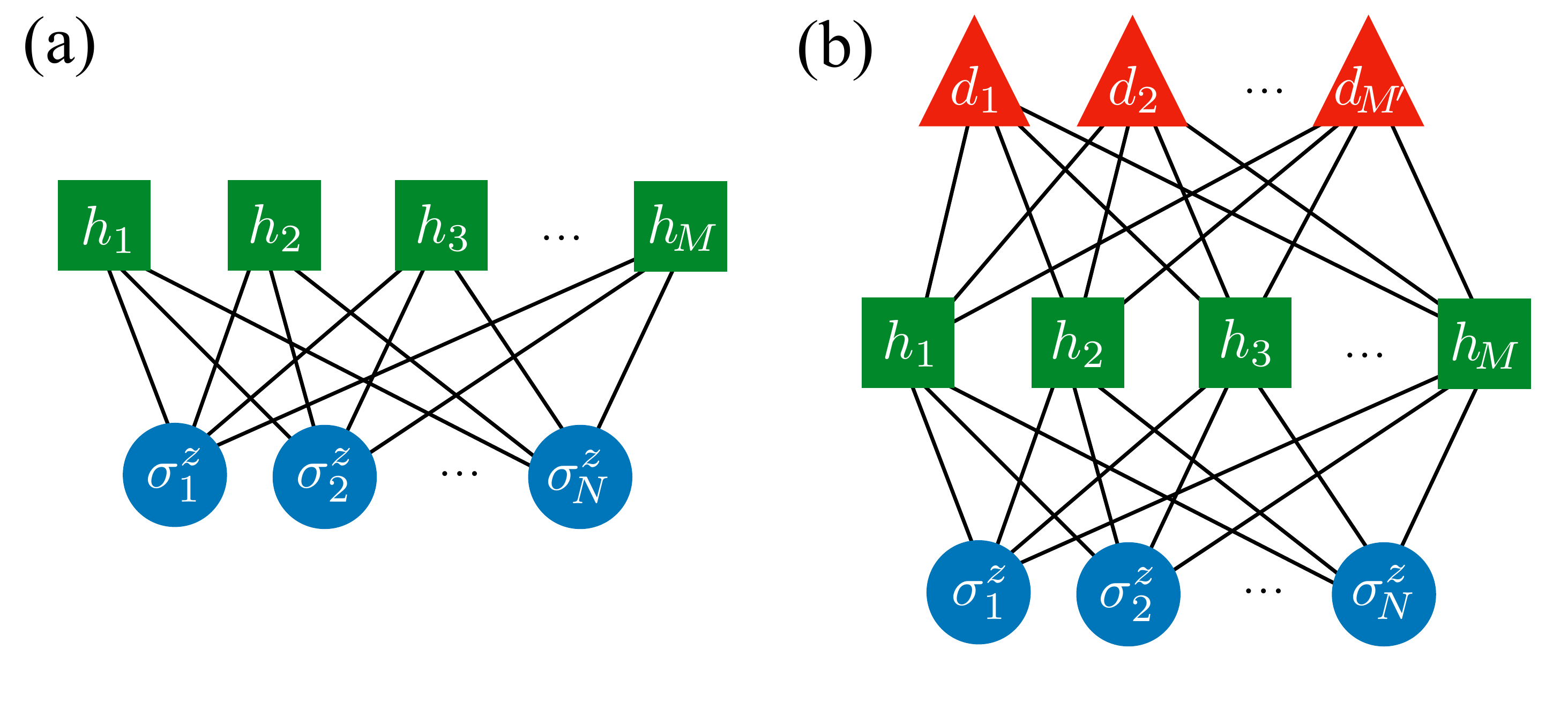}
\caption{
Boltzmann machines for representing quantum states of $S=1/2$ spin systems: 
(a) RBM and (b) DBM. 
}
\label{Fig_RBM_DBM}
\end{center}
\end{figure}

Here, we discuss several properties of Boltzmann machine wave functions using $S=1/2$ quantum spin systems.
By (i) identifying $\sigma_i^z$ with $v_i$ and (ii) extending the parameters to complex variables, Boltzmann machines can be applied to the representation of the quantum states of quantum spin systems (Fig.~\ref{Fig_RBM_DBM}).
In this case, the amplitude of the RBM wave function for a spin configuration $\sigma=(\sigma_i^z, \ldots \sigma_N^z)$ is given by 
\begin{eqnarray}
  \psi(\sigma)  \!  &=&  \! \!   \sum_h
  \exp  \Bigl( \sum_{i} a_i \sigma_i^z   + \sum_{i,j}  W_{ij} \sigma_i^z h_j  + \sum_{j} b_j h_j \Bigr)
  \nonumber  \\
   &=&  \!    \exp \Bigl( \sum_{i} a_i \sigma_i^z\Bigr) \! \times \!  \prod_j   2 \cosh \Bigl( b_j + \sum_{i} W_{ij}  \sigma_i^z \Bigr ).
   \label{eq_RBM_wf}
\end{eqnarray}
Hereafter, we omit the normalization factor for simplicity.
Similarly, the form of the DBM wave function reads 
\begin{eqnarray}
  \psi(\sigma)  =  \sum_{h,d} \exp  &\Bigl(&  \sum_{i} a_i \sigma_i^z  + \sum_{j} b_j h_j +  \sum_{k} b'_k d_k    \nonumber \\ 
                     && +    \sum_{i,j}  W_{ij} \sigma_i^z h_j  + \sum_{j,k} W'_{jk} h_j d_k \Bigr) .
    \label{eq_DBM_wf}
\end{eqnarray}

\subsubsection{Universal approximator \\}
\label{sec:universal}

As in the case of approximating probability distributions by the RBM with real parameters, 
the RBM with complex parameters also has the property of universal approximation for quantum state representation; in the limit of a large number of hidden spins, it can represent an arbitrary wave function with arbitrary accuracy~\cite{Clark_2018,Huang_2021}.  
The DBM has more flexible expressivity for quantum state approximation thanks to the existence of the deep layer~\cite{Gao_2017,Levine_2019}.

\subsubsection{Comparison to tensor networks: representability and entanglement entropy \\}
\label{Sec:EE}

The neural network method is complementary to the tensor network method~\cite{Verstraete_2008,Orus_2014}, including the density matrix renormalization group (DMRG)~\cite{White_1992,White_1993} using matrix product states (MPSs). 
This is because tensor networks are also known as universal approximators in the limit of large bond dimensions.

In the case of finite bond dimensions, the representability of tensor networks is well described by entanglement entropy. 
Tensor networks with finite bond dimensions obey the area law of entanglement entropy. 
For example, in the case of MPS, the maximum entanglement entropy that an MPS with a bond dimension of $\chi$ can exhibit is $\log \chi$. 
Therefore, tensor networks are good at approximating quantum states with area-law entanglement entropy. 

On the other hand, no such good indicator is known for Boltzmann machine wave functions (or, more generally, neural network wave functions) with finite hidden degrees of freedom. 
Therefore, there is no general theory (to the best of our knowledge) on how the approximation accuracy improves with the number of hidden degrees of freedom and how it asymptotically approaches the limit of universal approximation.

Nevertheless, several properties are known about the entanglement entropy of Boltzmann machine wave functions. 
For this purpose, we define two types of RBMs: long-range RBM and short-range RBM. 
In the former [latter], the hidden spins have finite coupling parameters to ${\mathcal O}(N)$ [${\mathcal O}(1)$] visible spins; thus, the coupling becomes long-ranged [short-ranged].
Then, we can show that quantum states represented by a ``long-range RBM''  can exhibit volume-law entanglement entropy~\cite{Deng_2017,Glasser_2018}.
In fact, the maximal volume-law entangled state in a system consisting of one-dimensional $N$ qubits ($S=1/2$ quantum spins) can be constructed by an RBM with ${\mathcal O}(N)$ complex parameters~\cite{Deng_2017}.
However, when the quantum state is represented by a short-range RBM, the entanglement entropy follows the area law~\cite{Deng_2017,Glasser_2018}.

\subsubsection{Comparison to tensor networks: transformation between MPS and Boltzmann machines \\}

Whether the neural network wave functions can be transformed into tensor network wave functions, or vice versa, is an interesting question to clarify the relationship between the two.
As an example, we shall consider the transformation rule between matrix product states and Boltzmann machine quantum states.

\paragraph{From RBM to MPS.}
Ref.~\cite{Chen_2018} discussed a conversion law from RBM to MPS. 
In the case of long-range RBM, analytic conversion requires an exponentially large bond dimension $\chi$ with respect to the system size.
However, when the coupling parameters of the RBM are local, the required bond dimension becomes smaller.
Please refer to Ref.~\cite{Chen_2018} for the specific transformation formulas. 
The magnitude of the required bond dimension can also be understood in terms of entanglement entropy.
As discussed in Sec.~\ref{Sec:EE}, quantum states represented by the RBM can exhibit volume-law entanglement entropy.
Because the maximum entanglement entropy of an MPS with bond dimension $\chi$ is $\log \chi$, 
an exponentially large bond dimension is required to represent quantum states with volume-law entanglement entropy.

\paragraph{From MPS to RBM/DBM.}
As for the inverse transformation, to the best of our knowledge, there is no general transformation rule from MPS to RBM.
However, owing to its universal approximation ability, the RBM can reproduce an arbitrary MPS in the limit of a large number of hidden spins.
On the other hand, in the case of the DBM having a higher representation power than the RBM, the number of hidden spins required for the transformation has been discussed~\cite{Gao_2017,Huang_2021}. 
In Ref.~\cite{Huang_2021}, it was shown that MPS with bond dimension $\chi$ for $N$ qubit systems has a DBM representation with ${\mathcal O}(N \chi^2)$ hidden degrees of freedom (hidden and deep spins).

\subsubsection{Comparison to tensor networks: short summary \\}

Because of the difference in the properties of entanglement entropy, quantum states with volume law entanglement entropy may be efficiently represented with artificial neural networks (for example, maximal volume-law entangled state in one-dimensional $N$ qubit systems, as discussed in Sec.~\ref{Sec:EE}), but not by tensor networks.

On the other hand, there are examples in which tensor networks have efficient representations with fewer parameters.
For example, whereas the Affleck-Kennedy-Lieb-Tasaki (AKLT)~\cite{Affleck_1987} state can be represented by an MPS with $\chi=2$~\cite{Orus_2014}, in the case of the RBM using $\sigma^z$ basis, a long-range RBM with nonlocal coupling parameters is needed to represent the hidden antiferromagnetic order in the AKLT state~\cite{Chen_2018}.
The number of parameters is ${\mathcal O}(N)$ in the case of MPS and increases to ${\mathcal O}(N^2)$ when RBMs are used~\cite{Glasser_2018} \footnote{More recently, it was shown that the RBM can represent the AKLT state efficiently with ${\mathcal O}(N)$ parameters by choosing the basis appropriately~\cite{Lu_2019,Pei_2021}.}. 
This example shows that the capacity of entanglement entropy is not the only factor that determines expressive power~\cite{Chen_2018}.

Thus, it is not the case that either tensor networks or artificial neural networks are always superior.
Rather, they are complementary.
Further discussion of the relationship between the two will lead to a better understanding of quantum state representation.

\section{RBM for zero temperature calculations}
\label{Sec_RBM_zero}

In the following, we will discuss methods using Boltzmann machines for zero-temperature and finite-temperature calculations. 
In this section, we introduce a variational method based on RBM for zero-temperature simulations.
This can be understood as an approximate (but accurate) representation of the ground state obtained by numerically optimizing the RBM parameters. 
In the next section (Sec.~\ref{Sec_DBM_zero}), taking advantage of the DBM's more flexible representability than the RBM, we introduce an analytic construction of the DBM that represents the ground state. 
As we will see in Sec.~\ref{Sec_DBM_zero}, this analytical construction connects a well-known physics concept, path-integral formalism, with a machine-learning concept, deep learning. 
This DBM method has the advantage of being exact except for the Suzuki-Trotter error; however, its applicability is limited by the negative sign problem as in the QMC method.
Therefore, for systems where the sign problem cannot be avoided, the former variational approach, which can avoid the explicit sign problem, will work better. 
Finally, we discuss an extension to finite-temperature simulations in Sec.~\ref{Sec_DBM_finite}.

\subsection{Variational method}

Among the eigenstates of the Hamiltonian, the ground state has the lowest energy and is the most stable quantum state at zero temperature. 
Obtaining an accurate representation of the ground state is a long-standing challenge not only in condensed matter physics but also in various fields, including particle physics, nuclear physics, and quantum chemistry. 

Here, we introduce a variational method using RBMs. 
Since the ground state is the lowest-energy state, the energy can be considered as a loss function. 
Thus, the problem of finding the ground state by minimizing energy can be viewed as a machine-learning task, i.e., optimization of a nonlinear function (RBM) using the nonlinear loss function (energy) in a high-dimensional parameter space~\cite{Melko_2019}.

\subsubsection{Comparison to conventional variational method}

Conventional variational methods often assume the form of a variational wave function based on some physical insights.
While this approach has the advantage of requiring a small number of parameters and being intuitively easy to understand, it is difficult to systematically improve its accuracy.
On the other hand, the artificial neural network method, in principle, can be systematically improved by increasing the size of the network (see Sec.~\ref{sec:universal}).
It may be possible to extract nontrivial quantum correlations even in situations where physical intuition is difficult to obtain.

\subsubsection{Optimization of RBM \\}

Here, we discuss how to optimize the RBM variational parameters to minimize energy (loss function).
Among various optimization methods, we introduce the stochastic reconfiguration (SR) method~\cite{Sorella_2001}. 
Some studies employ simpler methods such as stochastic gradient descent and its refined variants (e.g., Adam~\cite{Kingma_2014}).
We choose this method because it is closely related to the imaginary-time Hamiltonian evolution of quantum states, and hence the SR method greatly stabilizes the optimization~\footnote{
Note that $e^{-\tau {\mathcal H}} \ket{\psi^{(0)}}$ converges to the ground state as $\tau \rightarrow \infty$, 
as long as the initial state $\ket{\psi^{(0)}}$ has a nonzero overlap with the ground state.
}. 
Essentially the same optimization method is known as the natural gradient method~\cite{Amari_1992,Amari_1998} in the machine learning community. 
Both were developed independently, and their relationship has only recently been recognized.

The SR method makes use of information about the geometry of quantum states (Fubini-Study metric). 
Hence, we first introduce the Fubini-Study metric and move on to the explanation of the update rule of the SR method.
The Fubini-Study metric is a metric on the space of quantum states, whose form is given by
   \begin{eqnarray}
    \mathcal{F}[\ket{\psi}, \ket{\phi}] := \arccos \sqrt{\frac{\braket{\psi | \phi} \braket{\phi | \psi}}{ \braket{\psi | \psi}\braket{\phi | \phi} }}.
    \end{eqnarray}
 For normalized quantum states $\ket{\bar{\psi}} \! = \! \frac{ \ket{\psi} }{  \sqrt{ \braket{ \psi | \psi } } } $ and $\ket{\bar{\phi}} \! = \! \frac{ \ket{\phi} }{ \sqrt{ \braket{ \phi | \phi } } } $,  the Fubini-Study metric reads
  \begin{eqnarray}
 \mathcal{F} \left[\ket{\bar{\psi}}, \ket{\bar{\phi}} \right] = \arccos  \left(  |  \!   \braket{\bar{\psi} | \bar{\phi}}  \!   |  \right).
     \end{eqnarray}
This can be interpreted as the angle between the two complex vectors. 
Therefore, the Fubini-Study metric is occasionally called the quantum angle. 
Now, we consider the infinitesimal form of this metric; 
for a variational quantum state $\ket{ \psi_\theta}$ characterized by the parameter set $\theta$, it is given by
\begin{eqnarray}
   \mathcal{F}^2[\ket{\psi_{\theta+\delta \theta}}, \ket{\psi_\theta}] = \sum_{kl} S_{kl}(\theta) \delta{\theta_k} \delta{\theta_l}.
   \label{Eq:smat_def}
 \end{eqnarray}
$S_{kl} (\theta) =  {\rm Re} \bigl[ Q_{kl} (\theta) \bigr]$ is the Fubini-Study metric tensor and is given as the real part of so called quantum geometric tensor $Q_{kl} (\theta)$:
\begin{eqnarray}
    Q_{kl}(\theta) =  \frac{\braket{\partial_k \psi_\theta | \partial_l \psi_\theta}}{\braket{\psi_\theta | \psi_\theta}} - \frac{\braket {\partial_k \psi_\theta|  \psi_\theta} }{\braket{\psi_\theta | \psi_\theta}} \frac{\braket {\psi_\theta | \partial_l \psi_\theta}}{\braket{\psi_\theta| \psi_\theta}}.  
 \label{Eq:qmat_def}
 \end{eqnarray}
The Fubini-Study metric can be regarded as a quantum version of the Fisher information metric and thus has a close relationship with information geometry~\cite{Stokes_2020_quantum}.

As described above, the SR optimization is closely related to the imaginary-time evolution of quantum states. 
The update rule can be derived by trying to reproduce the infinitesimal imaginary-time evolution $e^{-2\delta_\tau\mathcal{H}}\ket{\psi_{\theta}}$ by the variational state with updated parameters $\ket{\psi_{\theta + \delta \theta}}$: 
\begin{eqnarray}
    \delta \widetilde{\theta} = \mathop{\rm arg~min}\limits_{\delta\theta} \mathcal{F}[e^{-2\delta_\tau\mathcal{H}}\ket{\psi_{\theta}}, \ket{\psi_{\theta + \delta \theta}}]
    \label{Eq_SR_goal}
\end{eqnarray}
After some algebra (see \ref{sec_SR_derivation} for details), we obtain the following update rule: 
 \begin{eqnarray}
    \delta \theta_k^{(t)} =  - \delta_\tau \sum_l  \bigl (S^{(t)} \bigr)^{-1}_{kl}  g_l^{(t)},
    \label{Eq:SR_update}
 \end{eqnarray}
where $t$ indicates the iteration number of optimization~\footnote{Replacing the $S$ matrix with the identity matrix (Hessian matrix) results in an update corresponding to the stochastic gradient descent method (Newton's method).}. 
$g_l^{(t)}$ is the gradient of the total energy (loss function) with respect to the variational parameters 
\begin{eqnarray}
  g_k^{(t)} = \left. \frac{ \partial E_\theta } { \partial \theta_k } \right |_{\theta = \theta^{(t)}}. 
  \label{Eq:gradient}
\end{eqnarray}
The move in Eq.~(\ref{Eq:SR_update}) can also be interpreted as the steepest descent direction with respect to the quantum information geometry~\cite{Stokes_2020_quantum}. 

The quantities used in the SR method can be estimated using the Monte Carlo method (see \ref{sec_SR_MC} for details).
Efficient techniques to compute the gradient with respect to the variational parameters, such as backpropagation~\cite{Rumelhart_1986} (reverse mode of automatic differentiation), can also be combined.

\subsubsection{Calculation procedure \\}
\label{sec_procedure}

Here, we summarize the calculation procedure by focusing on the variational approach for ground-state calculations.
\begin{enumerate}
\item  {\bf Initialization}: Construct a variational ansatz based on the RBM and initialize RBM variational parameters. We often put small random numbers. 
\item  {\bf Optimization}: Optimize the RBM parameters to minimize the loss function (total energy). In this paper, we employ the SR scheme for this purpose. 
\item  {\bf Post-process}: Once we obtain the optimized wave function, we can compute various physical quantities such as the correlation function, as well as the total energy. 
\end{enumerate}
When random numbers are used for the initialization of the parameters, multiple optimization calculations are often performed using different sets of initial parameters. Then, for the calculation of physical quantities in step (iii), we employ the lowest-energy state among the optimized variational states.

\begin{figure*}[tbp]
\vspace{0cm}
\begin{center}
\includegraphics[width=0.98\textwidth]{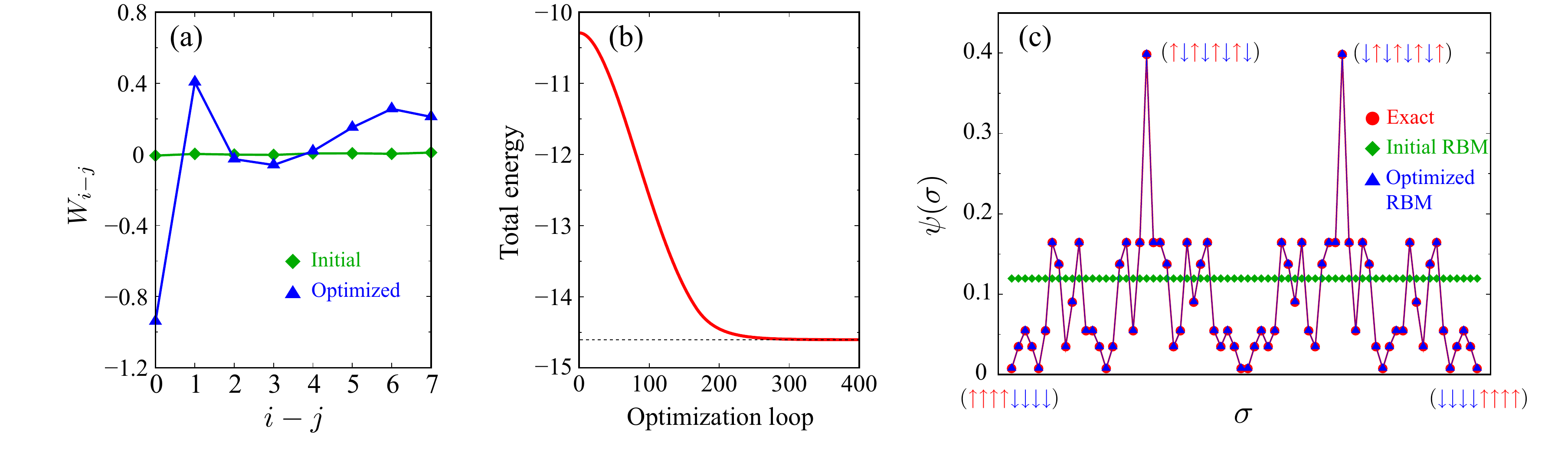}
\caption{
Demonstration of the variational method based on the RBM wave function. 
The method is applied to the one-dimensional Heisenberg model on the 8-site spin chain with the periodic boundary condition. 
(a) Initial and optimized RBM variational parameters in Eq.~(\ref{Eq.RBM_demo}).
Because of the translational symmetry, the origin of $i-j$ is arbitrary. We set the origin of $i-j$ so that $|W_0| $ is the largest among $|W_{i-j}|$. 
(b) Total energy (= loss function) as a function of the optimization loop. 
(c) Initial and optimized RBM wave functions and comparison to the exact diagonalization result. 
}
\label{Fig_demonstration}
\end{center}
\end{figure*}

\subsubsection{Demonstration \\}
\label{sec:demo}

Here, we demonstrate the application of the RBM wave function following the procedure described above. 
As an example, we consider the one-dimensional Heisenberg model on the 8-site spin chain with the periodic boundary condition. 
The Hamiltonian reads 
\begin{eqnarray}
 {\mathcal H} =  \sum_{i=1}^{8} ( - \sigma_i^x \sigma_{i+1}^x - \sigma_i^y \sigma_{i+1}^y + \sigma_i^z \sigma_{i+1}^z ).
\end{eqnarray}
This form is obtained by rotating the spin quantization axis by 180 degrees about the $z$ axis for one of the two sublattices (gauge transformation) from the usual form $ {\mathcal H} =  \sum_i (  \sigma_i^x \sigma_{i+1}^x + \sigma_i^y \sigma_{i+1}^y + \sigma_i^z \sigma_{i+1}^z )$.
With this form, the amplitude of the ground-state wave function becomes positive $\psi(\sigma) > 0$ for all $\sigma$ in the sector with $\sum_i \sigma_i^z = 0 $.

First, we construct a variational ansatz based on the RBM. 
The system has translational symmetry; therefore, we impose translational symmetry in the variational ansatz.
For this purpose, we impose translational symmetry on $W_{ij}$ parameters.
In addition, one way to preserve the symmetry between the up and down spins is to set the bias terms to zero, i.e., $a_i=0$ and $b_j =0$. 
When the number of hidden spins is equal to the number of physical spins, the RBM wave function reads  
\begin{eqnarray}
  \psi(\sigma)  =   \prod_{j=1}^8   2 \cosh \Bigl(  \sum_{i=1}^8 W_{i-j}  \sigma_i^z \Bigr ).
  \label{Eq.RBM_demo}
\end{eqnarray}
In the present case, the wave function is real and positive $\psi(\sigma) > 0$; thus, the $W_{i-j}$ parameters are taken to be real.
In this setup, the number of independent parameters is 8 ($W_{i-j}$ with $i-j = 0, \ldots, 7$).

As the initialization of the variational parameters, we put small random numbers [Fig.~\ref{Fig_demonstration}(a)]. 
We then optimize the variational parameters using the SR method to minimize the total energy (loss function) [Fig.~\ref{Fig_demonstration}(b)]. 
The optimized $W_{i-j}$ parameters show a sign change between $W_0$ where $|W_{i-j}|$ is largest and $W_1$ where $|W_{i-j}|$ is second largest, which indicates that the RBM learns the antiferromagnetic correlation [Fig.~\ref{Fig_demonstration}(a)]. 
For the present system size (8-site spin chain), we can easily perform exact diagonalization and list all wave function amplitudes in the sector with $\sum_i \sigma_i^z = 0 $.
While the initial RBM wave function has no recognizable features, the optimized RBM wave function reproduces the exact result quite well [Fig.~\ref{Fig_demonstration}(c)].
Once we obtain the optimized RBM wave function, we can compute, for example, the spin-spin correlation function (not shown).

When the system size becomes large, it is not possible to compute the wave function amplitudes for all the spin configurations. 
In this case, physical quantities are estimated numerically using the Monte Carlo method (see \ref{sec_SR_MC}).
If the number of hidden spins is polynomial with respect to $N$, the computational cost scales polynomially with respect to $N$.

\subsection{Results by Carleo and Troyer}
\label{Sec_Carleo_Troyer}

\begin{figure*}[tbp]
\vspace{0cm}
\begin{center}
\includegraphics[width=0.98\textwidth]{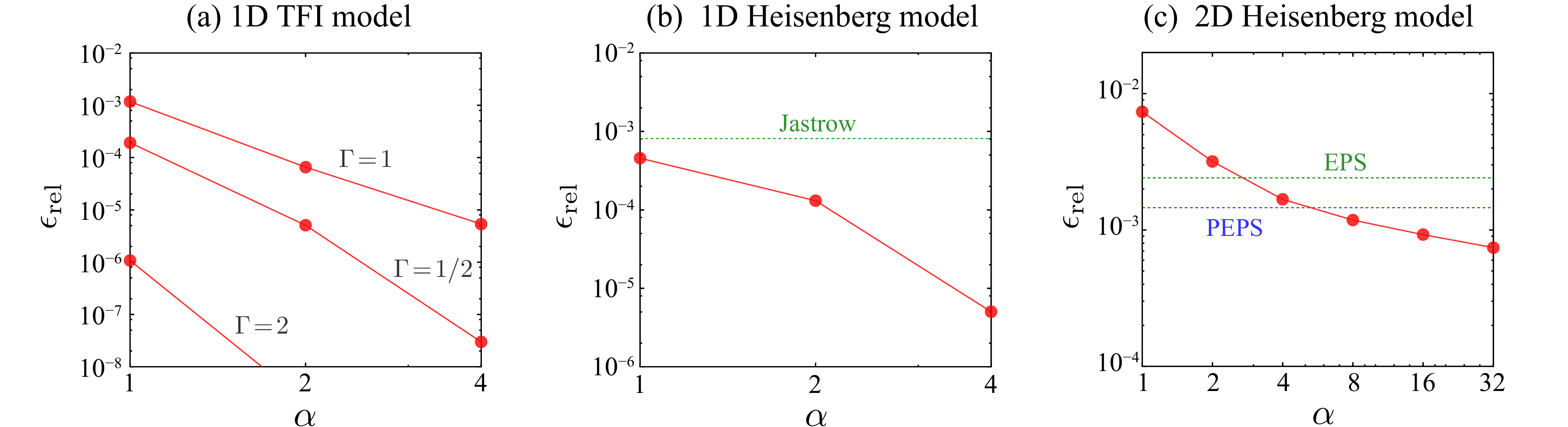}
\caption{
Results of variational calculations using RBM by Carleo and Troyer~\cite{Carleo_2017}.
Relative error $\epsilon_{\rm rel}$ of the RBM variational energy for 
(a) one-dimensional transverse-field Ising (TFI) model ${\mathcal H} = -\sum_{i}   \sigma_i^z \sigma^z_{i+1} - \Gamma \sum_i \sigma^x_i $ (periodic boundary condition, 80 sites) 
(b) one-dimensional Heisenberg model ${\mathcal H} =  \sum_{i} (  - \sigma_i^x \sigma^x_{i+1}   - \sigma_i^y \sigma^y_{i+1}  +  \sigma_i^z \sigma^z_{i+1})$ (periodic boundary condition, 80 sites), 
and (c) two-dimensional Heisenberg model ${\mathcal H} =  \sum_{\langle i,j \rangle } (  - \sigma_i^x \sigma^x_{j}   - \sigma_i^y \sigma^y_{j}  +  \sigma_i^z \sigma^z_{j})$ (periodic boundary condition, $10\times10$ square lattice).
$\alpha$ ($=M/N$) is a parameter that controls the accuracy of the RBM.
$\epsilon_{\rm rel}$ is given by $\epsilon_{\rm rel} = ( E_{\rm RBM}(\alpha) - E_{\rm exact} )/ | E_{\rm exact} | $.
``Jastrow'' in (b) shows the result of Jastrow ansatz, and EPS and PEPS in (c) are the results of tensor networks taken from Refs.~\cite{Mezzacapo_2009} and  \cite{Lubasch_2014}, respectively. 
}
\label{Fig_results_Carleo}
\end{center}
\end{figure*}

The variational method based on the RBM was first introduced by Carleo and Troyer~\cite{Carleo_2017}.
They applied the RBM wave function to the one-dimensional transverse-field Ising model, the one-dimensional Heisenberg model, and the two-dimensional Heisenberg model on a square lattice~\cite{Carleo_2017}.
Since these models have no geometrical frustration, the values of the ground-state wave function can be made positive and real by a proper gauge transformation (the gauge transformation was also used in Sec.~\ref{sec:demo}).
Using this fact, Carleo and Troyer also set the variational parameters to real numbers, as in Sec.~\ref{sec:demo}.
They also imposed the same translational symmetry on $W$ parameters (so that the number of hidden spins $M$ is an integer multiple of the number of visible spins $N$).
Defining $\alpha = M/N$, $\alpha$ is the parameter that controls the accuracy of the RBM wave function ($\alpha=1$ was used in the demonstration in Sec.~\ref{sec:demo}).

When introducing a new method, it is essential to verify its accuracy. 
Fig.~\ref{Fig_results_Carleo} shows the benchmark results comparing the RBM variational energies and numerically exact energies. 
We see that, for all Hamiltonians, increasing $\alpha$ improves the accuracy of the RBM variational state and reduces the relative error in energy, as expected. 
The results for the two-dimensional Heisenberg model shown in Fig.~\ref{Fig_results_Carleo}(c) show that the RBM can achieve better accuracy than those obtained from tensor networks such as entangled-plaquette states (EPSs)~\cite{Gendiar_2002,Mezzacapo_2009} and projected entangled pair states (PEPSs)~\cite{Verstraete_2004_arXiv}\footnote{The comparison was made under the periodic boundary condition. We note that the performance of tensor networks is much better under the open boundary condition.}.

\subsection{Extensions to strongly-correlated Hamiltonians}

The application of the variational method based on artificial neural networks started from frustration-free spin systems as introduced above (Sec~\ref{Sec_Carleo_Troyer}). 
When geometrical frustration is absent, the QMC method can be applied without the negative sign problem (see also Sec.~\ref{Sec_introduction}). 

However, such cases are special, and general quantum many-body Hamiltonians exhibit unavoidable negative sign problems. 
Constructing a good ground-state variational hypothesis for such problems is an important task for variational methods.
Here, among many challenging unsolved quantum many-body Hamiltonians, we consider fermion and frustrated spin systems as examples.

\subsubsection{Fermion systems\\}
\label{sec_fermion}

\begin{table*}[tb]
\caption{
Applications to fermion systems.
ANN stands for artificial neural network. 
When referring to molecules as a target, the target also includes, e.g., atoms, diatoms, and hydrogen chains. 
}
\label{table:fermion}
\centering
\vspace{0.3cm}
\begin{tabular}{@{\    }  l @{\  \  \   \  \  \ }  l @{\  \  \  \  \  \  }  l @{\  \  \ \ \ \ }   l  }
\hline \hline
 Work                                 &  Method   &  Main target               &  Ref.    \\ \hline
Nomura {\it et al.} (2017)   &  Jastrow-type   &  lattice model     &  \cite{Nomura_2017}    \\ 
Luo {\it et al.} (2019)          & backflow-type       & lattice model      & \cite{Luo_2019}  \\
Han {\it et al.} (2019)         &  anti-symmetrized ANN   &  continuous-space, molecules     &  \cite{Han_2019}    \\ 
Choo {\it et al.} (2020)        &  mapping to spin model &  continuous-space, molecules     &  \cite{Choo_2020}    \\ 
Pfau {\it et al.} (2020)         &  backflow-type   &  continuous-space, molecules     &  \cite{Pfau_2020}    \\ 
Hermann {\it et al.} (2020)   &  backflow+Jastrow-type   &  continuous-space, molecules     &  \cite{Hermann_2020}    \\ 
Stokes {\it et al.} (2020)      & Jastrow-type   &  lattice model     &  \cite{Stokes_2020}    \\ 
Yoshioka {\it et al.} (2021)   & mapping to spin model &  continuous-space, periodic     &  \cite{Yoshioka_2021}    \\ 
Inui {\it et al.} (2021)           &  anti-symmetrized ANN    &  lattice model     &  \cite{Inui_2021}    \\ 
Moreno {\it et al.} (2022)     &  augmented Hilbert space+projection    &  lattice model     &  \cite{Moreno_2022}    \\ 
Cassella {\it et al.} (2023)   &  backflow-type  & continuous-space, periodic     & \cite{Cassella_2023} \\
 \hline \hline
\end{tabular}
\end{table*}

Analyzing the electronic structures of strongly-correlated materials is one of the most fundamental problems in physics. 
Because electrons are fermionic particles, it is highly important to extend the method to fermionic systems. 

When we consider applications to fermion systems, we must consider the anti-commutation relations of fermion operators. 
Studies on fermion systems can be classified into two types (Table~\ref{table:fermion}): 
One takes account of the anti-symmetric property in wave functions~\cite{Han_2019,Inui_2021,Nomura_2017,Stokes_2020,Luo_2019,Pfau_2020,Hermann_2020,Cassella_2023,Moreno_2022}, while the other incorporates it in Hamiltonians~\cite{Choo_2020,Yoshioka_2021}.
Among studies, Refs.~\cite{Nomura_2017}, \cite{Choo_2020}, and \cite{Yoshioka_2021} employed RBMs in a variational ansatz, with Ref.~\cite{Nomura_2017} using the former approach and Refs. \cite{Choo_2020} and \cite{Yoshioka_2021} using the latter.

\begin{figure}[tbp]
\vspace{0cm}
\begin{center}
\includegraphics[width=0.49\textwidth]{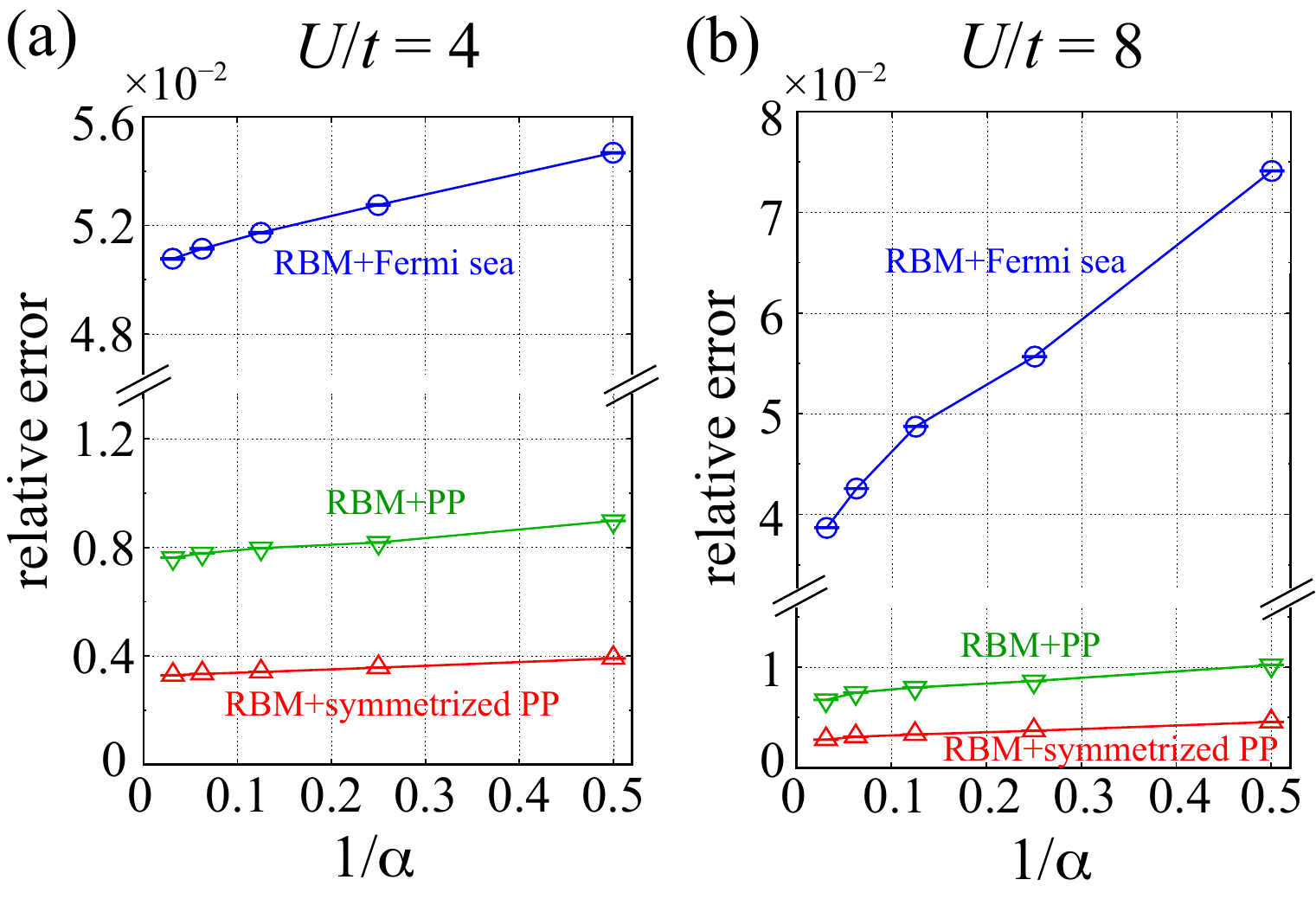}
\caption{
Relative errors of variational energies for the Hubbard model on the $8\times 8$ square lattice (periodic and anti-periodic boundary conditions) at (a) $U/t=4$ and (b) $U/t=8$. 
Numerically exact references are obtained by the QMC method. 
$\alpha$ ($=M/N$) is a parameter that controls the accuracy of the RBM part.
Adapted with permission from Ref.~\cite{Nomura_2017}. Copyright 2017 by the American Physical Society.
}
\label{Fig_results_Nomura_2017}
\end{center}
\end{figure}

In Ref.~\cite{Nomura_2017}, the RBM is combined with the pair-product (PP) state~\cite{Tahara_2008}.
The pair-product wave function is also known as the geminal wave function in quantum chemistry and satisfies the anti-symmetric property. 
In this case, the RBM is employed as a generalization of correlation factors such as the Gutzwiller~\cite{Gutzwiller_1963} and Jastrow~\cite{Jastrow_1955} factors. 
Alternatively, we can reformulate the problem as approximating $\psi_{\rm exact} / \psi_{\rm PP}$ using RBM.
Therefore, when the combined wave function becomes sophisticated, it becomes easy for the RBM to learn the ground state. 
This trend was confirmed by the benchmark results for the Hubbard model on the $8\times8$ square lattice (Fig.~\ref{Fig_results_Nomura_2017}). 
See also Sec.~\ref{sec_frustrated_systems} for a discussion of the possible importance of the combination.

In Refs.~\cite{Choo_2020} and \cite{Yoshioka_2021}, fermionic Hamiltonians are first mapped onto quantum spin Hamiltonians. 
For this transformation, one can employ, e.g., the Jordan-Wigner~\cite{Jordan_1928} or Bravyi-Kitaev~\cite{Bravyi_2002} transformations.
The RBM is then employed to approximate the ground state (excited states) of the mapped quantum-spin Hamiltonians.

\subsubsection{Frustrated quantum-spin systems \\}
\label{sec_frustrated_systems}

Quantum spin systems with geometrical frustration are fascinating systems that may host an exotic state of matter, such as quantum spin liquid~\cite{Balents_2010,Y_Zhou_2017}. 
However, at the same time, it is numerically challenging to analyze frustrated quantum spin systems. 
Extensions to frustrated quantum spin systems are one of the major topics in the field of neural-network quantum states.

\paragraph{Benchmarks \\}

Among various frustrated quantum-spin Hamiltonians, the $J_1$--$J_2$ Heisenberg model on square lattices has been most intensively studied. 
Therefore, we focus on the $J_1$--$J_2$ Heisenberg model here.

\begin{table}[tb]
\caption{
Comparison of ground-state energy among various variational methods for the $J_1$--$J_2$ Heisenberg model with $J_2/J_1=0.5$ on the $10\times10$ square lattice (periodic boundary condition).  
The wave functions in bold font employ artificial neural networks.
Ref.~\cite{Hu_2013} applies $p$-th order Lanczos step on top of the optimized VMC state. 
CNN and GCNN stand for convolutional neural network and group convolutional neural network, respectively. 
}
\label{table:ene_comparion_10x10}
\centering
\vspace{0.3cm}
\begin{tabular}{@{\   }  l @{\  \  \   }  l @{\  \ }  l  }
\hline \hline
 Energy per site  &  Wave function &  Ref.    \\ \hline
$-0.49516(1)$    &  {\bf CNN}  &  \  \cite{Choo_2019} \\
$-0.49521(1)$    &  VMC($p$=0) &  \  \cite{Hu_2013} \\
$-0.495530$       &  DMRG & \   \cite{Gong_2014} \\
$-0.49575(3)$    &  {\bf RBM-fermionic w.f.}&\   \cite{Ferrari_2019} \\
$-0.49717$    &  {\bf CNN}  &  \ \cite{M_Li_2022} \\
$-0.497437(7)$    &  {\bf GCNN}  &  \ \cite{Roth_2023} \\
$-0.497549(2)$  &  VMC($p$=2)  &\   \cite{Hu_2013} \\
$-0.497629(1)$  &  {\bf RBM+PP} & \  \cite{Nomura_2021_PRX} \\
\hline \hline
\end{tabular}
\end{table}

\begin{figure}[tbp]
\vspace{0cm}
\begin{center}
\includegraphics[width=0.48\textwidth]{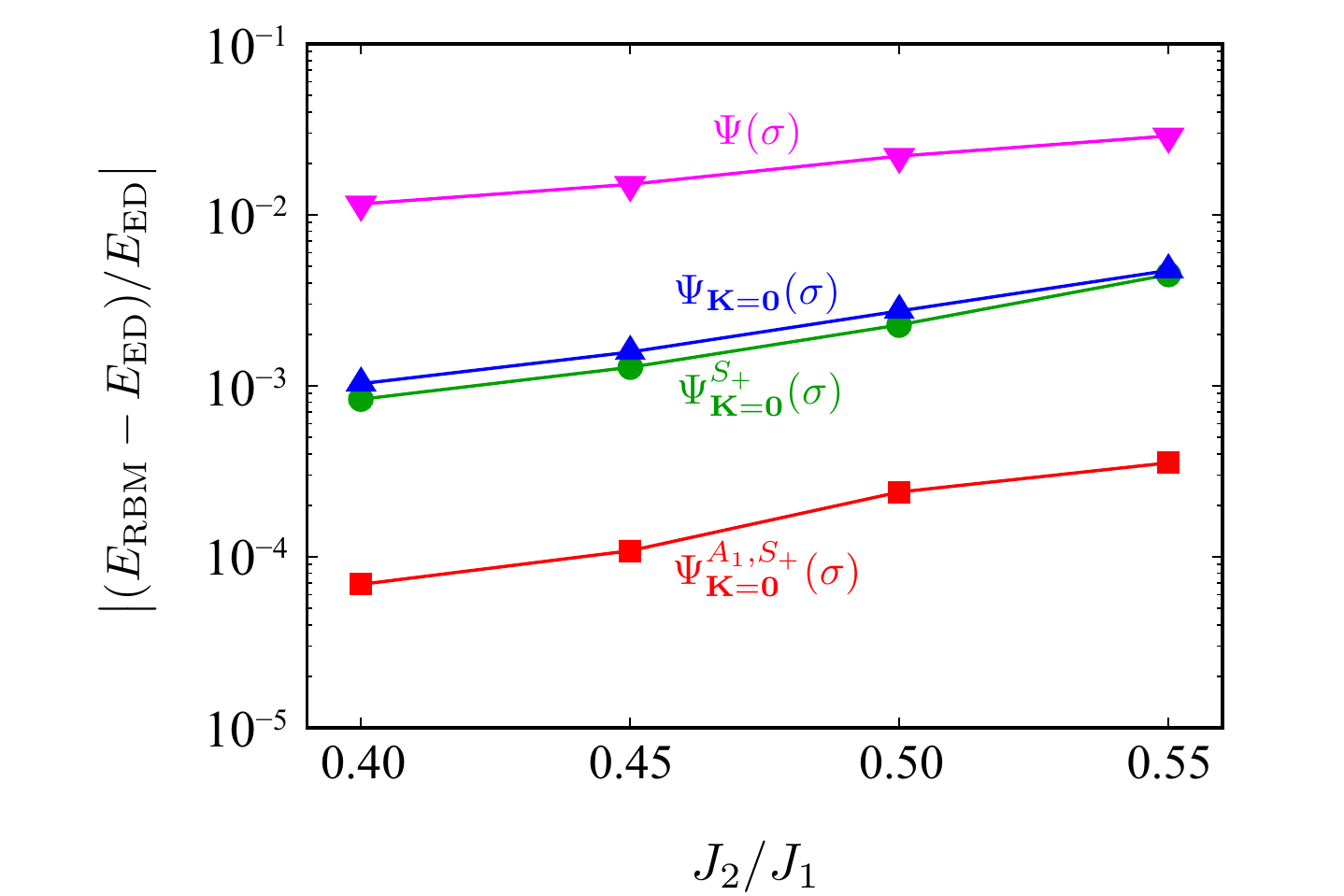}
\caption{
Effects of symmetrization on the accuracy of RBM wave function with $\alpha=M/N=2$ for the $J_1$--$J_2$ Heisenberg model on the $6\times6$ square lattice (periodic boundary condition).  
$E_{\rm ED}$ is the ground-sate energy obtained by exact diagonalization. 
Compared to the results of $\Psi(\sigma)$ without any symmetrization, 
the results with quantum-number projections show much better accuracy. 
Applied quantum-number projections are as follows: 
total momentum (${\bf K} = {\bf 0}$), spin-parity (even parity, represented as $S_+$), and $A_1$ irreducible representation of the $C_{4v}$ point group.
Reproduced from Ref.~\cite{Nomura_2021_JPCM} (\copyright \ 2021 IOP Publishing Ltd) by permission of IOP Publishing. All rights reserved.
}
\label{Fig_results_Nomura_2021}
\end{center}
\end{figure}

Table~\ref{table:ene_comparion_10x10} shows a comparison of the ground-state energy among various variational methods for the $J_1$--$J_2$ Heisenberg model with $J_2/J_1=0.5$ on the $10\times10$ square lattice (periodic boundary condition).
We leave several remarks from this comparison. 
\begin{itemize}
\item {\bf Symmetry.} 
To achieve high accuracy, it is important to employ symmetrization. 
Indeed, all highly accurate neural-network variational energies were obtained by a certain symmetrization of the wave function. 
Symmetrization can be achieved in several ways.
For example, Ref.~\cite{Carleo_2017}, which introduced the RBM wave function, employed the symmetrization of variational parameters. 
Recently, a projection onto a particular quantum number sector~\cite{Nomura_2021_JPCM} has often been employed (see Fig.~\ref{Fig_results_Nomura_2021} for how the accuracy is improved by quantum-number projections). 
Unfortunately, the performance depends on the symmetrization technique (see, e.g., Ref.~\cite{Reh_2023}); therefore, careful consideration of the symmetrization is necessary to obtain a good performance~\footnote{Another advantage of symmetrization is that excited states can be calculated by considering symmetry sectors that are different from those of the ground state. }.
\item {\bf Combination.} 
We can think of combining different ansatzs to construct a single variational state. 
This is practically important because there exists a tradeoff between trainability and representability in neural-network ansatzs.
The combination helps to reduce the number of parameters in the neural network part and mitigates the problem of trainability (see also the discussion related to Fig.~\ref{Fig_results_Nomura_2017} in Sec~\ref{sec_fermion}).  
Indeed, for $10\times10$ square lattice at $J_2/J_1=0.5$, the best accuracy among the compared methods is obtained by the combination of RBM and pair-product (PP) state that is mapped onto the spin Hilbert space~\cite{Nomura_2021_PRX}.
\item {\bf Trainability.} 
Although deep neural network architectures should be better than shallow architectures to obtain good variational results, the problem of trainability often causes problems in achieving good performance.
Introducing huge computational resources is a brute-force way to mitigate this problem, obtaining better accuracy using CNN compared to early results~\cite{M_Li_2022}. 
A careful symmetric construction also helps~\cite{Roth_2023}.
Improving the optimization algorithm is also an important issue (see, e.g., Ref.~\cite{Chen_arXiv} for the latest attempts).  
\end{itemize}

\paragraph{Beyond benchmarks \\}

Once a novel accurate variational method is developed, it is important to apply it to study the physics of unsolved quantum many-body Hamiltonians. 
Recently, such applications have started to emerge. 
For example, Ref.~\cite{Nomura_2021_PRX} applied the RBM+PP wave function to the $J_1$--$J_2$ Heisenberg model on a square lattice and revealed a quantum spin liquid phase. 
Ref.~\cite{Astrakhantsev_2021} employed RBM, CNN, and many-variable variational Monte Carlo (mVMC)~\cite{Misawa_2019} methods to investigate the Heisenberg model on a pyrochlore lattice and concluded that 
symmetry-broken state showing long-range dimer order is realized as a ground state instead of quantum spin liquid without any long-range order. 
As such, the phase of research is shifting from development/benchmarking to ``true applications'' (of course, the development of methods must continue at the same time).

\section{DBM for zero-temperature calculations}
\label{Sec_DBM_zero}

Here, we describe a quantum state representation using DBMs, connecting deep learning and a well-known physics concept, path-integral formalism. 
In the previous section (Sec.~\ref{Sec_RBM_zero}), to represent the ground states of quantum many-body systems with high accuracy, we described the numerical optimization of the RBM parameters using the SR method to reproduce the imaginary-time Hamiltonian evolution as accurately as possible within the representability of the RBM variational ansatz.
Compared to the shallow RBM, the DBM has more flexible representation capability thanks to the existence of the deep layer (Fig.~\ref{Fig_RBM_DBM})~\cite{Gao_2017,Levine_2019}. 
Ref.~\cite{Carleo_2018} focused on the improved representability and showed that the imaginary-time Hamiltonian evolution can be reproduced with arbitrary precision within the DBM framework (see also Ref.~\cite{Freitas_2018} for a similar attempt).  
This leads to an analytic construction of the DBM representation of ground states.

In more detail, the procedure of this method is as follows: 
First, the initial state is prepared as a DBM. 
We apply the imaginary-time Hamiltonian evolution with the Suzuki-Trotter decomposition 
$e^{- \tau {\mathcal H} } \simeq  \left (e^{- \delta_\tau {\mathcal H}_1 }  e^{ - \delta_\tau  {\mathcal H}_2 } \right)^{N_\tau}$ (${\mathcal H} = {\mathcal H}_1 + {\mathcal H}_2 $, $\delta_\tau = \tau /N_\tau $) to the DBM state. 
Each short-imaginary-time evolution can be reproduced exactly by changing the DBM structure (an analytical expression for the parameter can be derived).
In the limit of long imaginary-time evolution, we can obtain the DBM representation of the ground state as long as the initial DBM state is not orthogonal to the ground state.  
Once the analytical DBM representation of the ground state is obtained, physical quantities can be computed using the Monte Carlo method. 
Since there is no efficient way to trace out the hidden and deep spins analytically, for this purpose, Monte Carlo sampling must be performed on the hidden or deep spin (or both) degrees of freedom in addition to the visible spins (Sec.~\ref{sec_DBM_explanation}).

\begin{figure}[tbp]
\vspace{0cm}
\begin{center}
\includegraphics[width=0.48\textwidth]{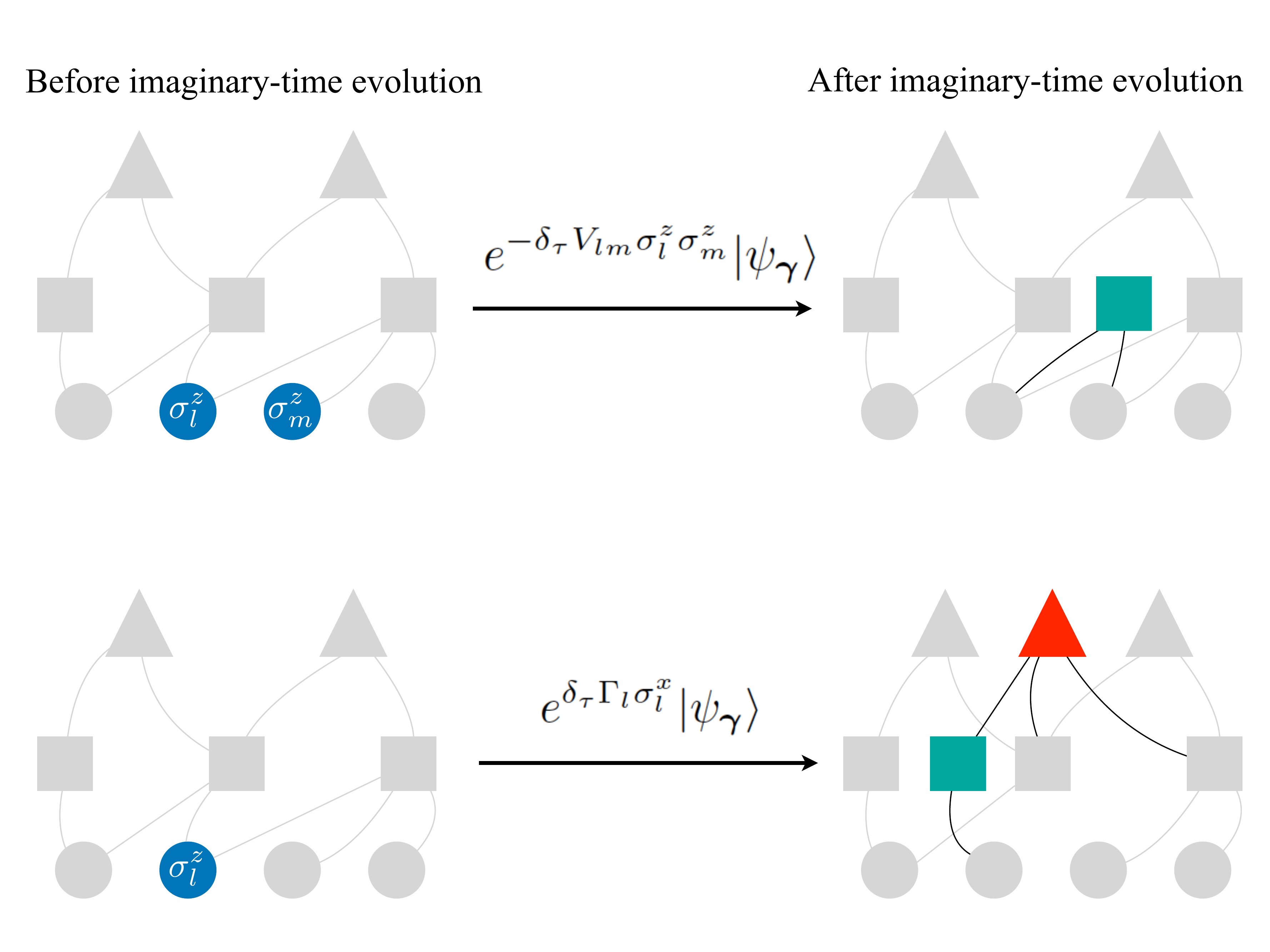}
\caption{
Evolution of DBM quantum state by short imaginary-time propagators, 
$e^{-{\mathcal H}_1 \delta_\tau }$ and $e^{-{\mathcal H}_2 \delta_\tau }$, of the transverse-field Ising model: 
${\mathcal H} = {\mathcal H}_1 + {\mathcal H}_2$, ${\mathcal H}_1 = \sum_{l<m} V_{lm} \sigma_l^z \sigma_m^z$, ${\mathcal H}_2 = - \sum_{l} \Gamma_l \sigma_l^x$. 
The relevant visible, hidden, and deep spins are highlighted in color. 
}
\label{Fig_DBM_zero}
\end{center}
\end{figure}

\begin{table*}[tb]
\caption{
Comparison between the RBM and DBM methods described in Sec.~\ref{Sec_RBM_zero} ad Sec.~\ref{Sec_DBM_zero}, respectively.
}
\label{table:RBM_DBM}
\centering
\vspace{0.3cm}
\begin{tabular}{ l @{\  \  \  \  \ }  l @{\  \  \  \  \ \  }    l  }
\hline \hline
                                         &  RBM (variational method) &  DBM  (quantum-to-classical mapping)            \\ \hline
Optimization and accuracy        &  numerical and approximate    &  analytical and exact  (up to Trotter error)     \\ 
Number of hidden units       &  smaller compared to DBM      &  $\propto$ (imaginary time)$\times$(system size)   \\ 
Sampling objects                 &   visible spins     & visible spins + hidden and/or deep spins  \\
Negative sign problem        &   applicable to frustrated systems    &  unavoidable in frustrated case \\ 
 \hline \hline
 \end{tabular}
\end{table*}

As an example, Fig.~\ref{Fig_DBM_zero} shows how the DBM can exactly reproduce the short imaginary-time evolution of the transverse-field Ising model.
To reproduce the evolution, new hidden and deep spins are introduced. 
Therefore, the size of the DBM network grows as a result of imaginary-time evolution;  
the number of hidden and deep spins is proportional to the system size and length of the imaginary time.
The benchmark calculations in Ref.~\cite{Carleo_2018} for several quantum-spin models confirmed that the constructed DBM indeed reproduces the ground state of each model. 

Finally, we briefly compare this DBM method with the RBM variational method (see Table~\ref{table:RBM_DBM}). 
This DBM method provides a way to perform quantum-to-classical mapping (note that DBMs are based on a classical spin model), and in a special case, the mapping becomes equivalent to path-integral formalism~\footnote{
When the mapping becomes equivalent to the path-integral formalism, deep spins play a role as auxiliary spins along imaginary time slices, and hidden spins mediate the interaction between the deep spins.  For more details, please refer to Ref.~\cite{Carleo_2018}.
}. 
While it has the advantage of not requiring cumbersome numerical optimization of parameters, unfortunately, the negative sign problem can not be avoided for, e.g., frustrated quantum-spin systems, as in the path-integral formalism;  
one can construct a DBM representation of the ground state, but negative Monte Carlo weights appear in the computation of physical quantities. 
Nevertheless, this method is of great significance in that it connects deep learning with the well-known physics concept and demonstrates the remarkable representative power of deep neural networks.

\section{DBM for finite-temperature calculations}
\label{Sec_DBM_finite}

So far, we have discussed calculations at zero temperature, where the quantum effect governs the property.  
An extension to finite temperature calculations is challenging because one needs to take into account thermal and quantum fluctuations simultaneously. 
Finite-temperature simulations are also important for comparison with experiments (note that experiments are performed at finite temperatures).

Here we describe two approaches for performing finite-temperature simulations using DBM~\cite{Nomura_2021_PRL}.
Both methods use the concept of ``purification'', in which a mixed state of a system is expressed in terms of a pure state of an extended system.

As a concrete example, we consider a quantum spin-1/2 system with $N$ spins. 
Considering an extended system with $N$ system spins and $N$ ancilla spins, the infinite-temperature state of the original system can be expressed as the following pure state of the extended system:
$ | \Psi(T \! =\! \infty) \rangle =  \bigotimes_{i=1}^{N} \left( | \uparrow \downarrow^\prime \rangle  +  | \downarrow \uparrow^\prime  \rangle \right)_i $
~\footnote{A pure state that reproduces the infinite-temperature state is not unique. Here, we chose one of the simplest forms.}.
From this state, the imaginary-time evolution by the Hamiltonian of the system leads to a pure state of the extended system corresponding to the thermal mixed state of the original system.
More explicitly, the pure state at temperature $T$ reads $ \ket{ \Psi (T) } = e^{-\beta \mathcal{H}/2}\otimes \mathbbm{1}' \ket{\Psi(T=\infty)}$ ($\beta=1/T$).
When the ancilla spin degrees of freedom are traced out, it gives the density matrix of the thermal equilibrium state.

\begin{figure}[tbp]
\vspace{0cm}
\begin{center}
\includegraphics[width=0.48\textwidth]{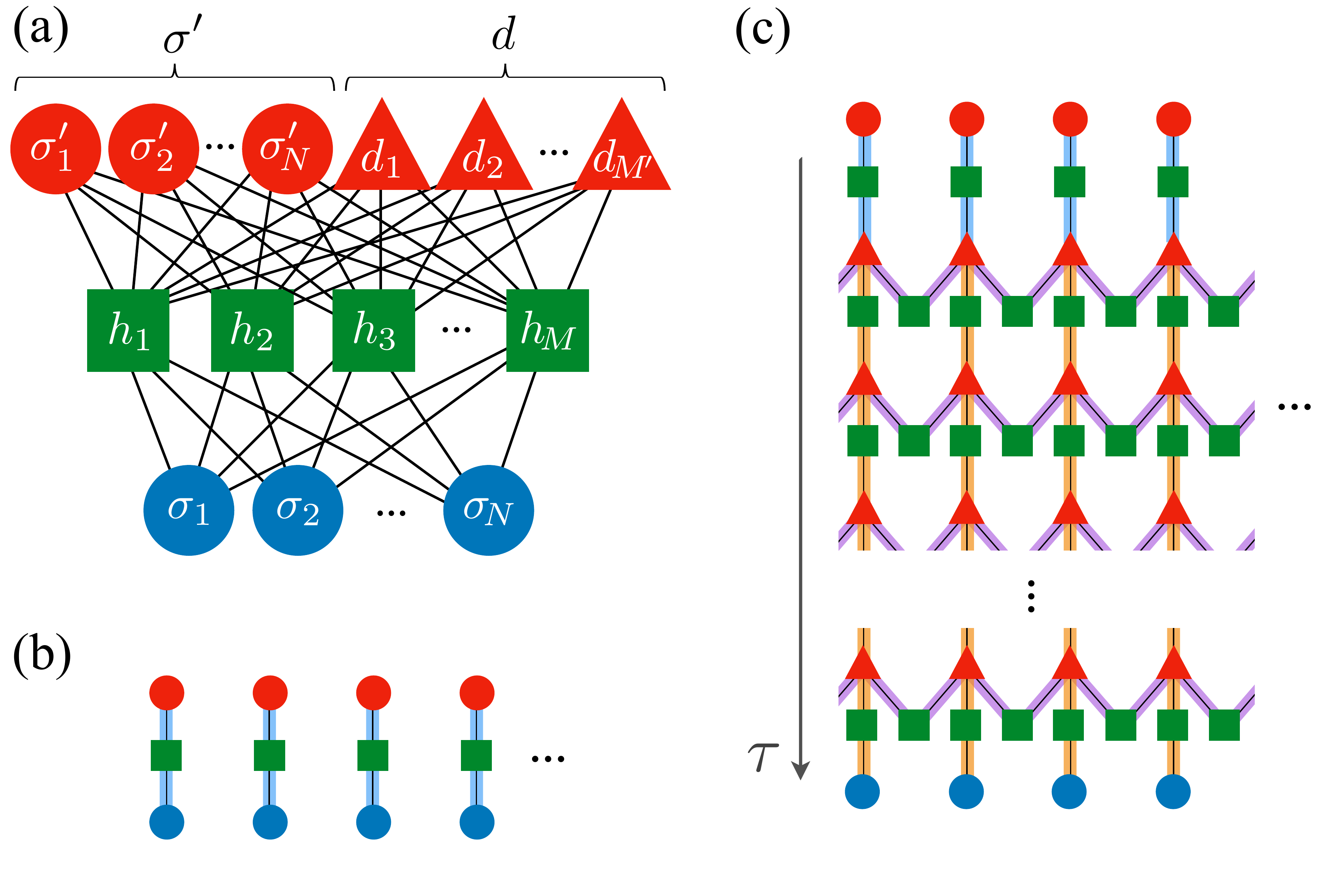}
\caption{
(a) Structure of DBM to represent pure states of the extended system during finite temperature calculations.
(b) DBM representing the pure state of the extended system $ | \Psi(T \! =\! \infty) \rangle $ corresponding to the infinite temperature of a quantum spin-1/2 system. 
(c) DBM representing the pure state of the extended system corresponding to the thermal equilibrium state of the one-dimensional transverse-field Ising model ${\mathcal H} = -J \sum_{i}   \sigma_i^z \sigma^z_{i+1} - \Gamma \sum_i \sigma^x_i $.
As described in Fig.~\ref{Fig_structure_BM}(d), the structure in (c) can be redrawn as the structure in (a). 
The magnitudes of the light blue, orange, and purple bonds in (b) and (c) are given by $\mathrm{i} \frac{\pi}{4}$, $\frac{1}{2} {\rm arcosh} \left( \frac{1}{{ \rm tanh} ( \Gamma \delta_\tau )}\right ) $, $\frac{1}{2}{\rm arcosh} (e^{2 J \delta_\tau })$, respectively.
Adapted with permission from Ref.~\cite{Nomura_2021_PRL}. Copyright 2021 by the American Physical Society.
}
\label{Fig_DBM_finite_T}
\end{center}
\end{figure}

Interestingly, the finite-temperature pure state $\ket{\Psi (T)}$ can be reproduced by the DBM shown in Fig.~\ref{Fig_DBM_finite_T}(a).
First, it can be shown analytically that the infinite-temperature pure state $ | \Psi(T \! =\! \infty) \rangle$ can be reproduced using the DBM shown in Fig.~\ref{Fig_DBM_finite_T}(b) (the corresponding DBM wave function reads $\Psi(\sigma,\sigma')   =   \prod_i 2 \cosh  \left[ \mathrm{i} \frac{\pi}{4} ( \sigma_i +  \sigma^{\prime}_{i} )  \right ]$).
The imaginary-time evolution from this state can be reproduced either (i) analytically as described in Sec.~\ref{Sec_DBM_zero} or (ii) numerically and approximately using the SR method described in Sec.~\ref{Sec_RBM_zero}.

In Method (i), the imaginary-time evolution is decomposed using Suzuki-Trotter decomposition as in Sec.~\ref{Sec_DBM_zero}, and each short-time evolution is reproduced exactly and analytically within the DBM framework.
As an example, we consider the one-dimensional transverse-field Ising model.
Following the procedure shown in Fig.~\ref{Fig_DBM_zero}, we can reproduce the imaginary-time evolution applied to the infinite-temperature DBM state  $ | \Psi(T \! = \! \infty) \rangle$ shown in Fig.~\ref{Fig_DBM_finite_T}(b). 
Fig.~\ref{Fig_DBM_finite_T}(c) shows the constructed DBM structure representing $\ket{\Psi (T)}$. 
Using this DBM structure, we can calculate physical quantities at finite temperatures by Monte Carlo sampling over configurations of physical spins $\sigma$, ancilla spins $\sigma'$, hidden spins $h$, and deep spins $d$.

In Method (ii), we consider the DBM structure without deep spins $d$ [i.e., we omit $d$ spins from the structure shown in Fig.~\ref{Fig_DBM_finite_T}(a)].
In this case, the DBM wave function describing the pure state of the extended system can be written analytically as $\Psi(\sigma,\sigma') \! =  \!  \prod_j 2 \cosh  \bigl[ b_j  +   \sum_i ( W_{ji}  \sigma_i   +   W'_{ji}  \sigma_i^{\prime} )  \bigr ]$.
It has a form similar to that of Eq.~(\ref{eq_RBM_wf}) in Sec.~\ref{Sec_BM_WF}, and this analytical form can be used as a variational ansatz. 
By numerically optimizing the DBM parameters using the SR method as described in Sec.~\ref{Sec_RBM_zero}, the imaginary-time evolution can be approximated as accurately as possible within the expressive capabilities of the variational wave function.

\begin{figure}[tbp]
\vspace{0cm}
\begin{center}
\includegraphics[width=0.48\textwidth]{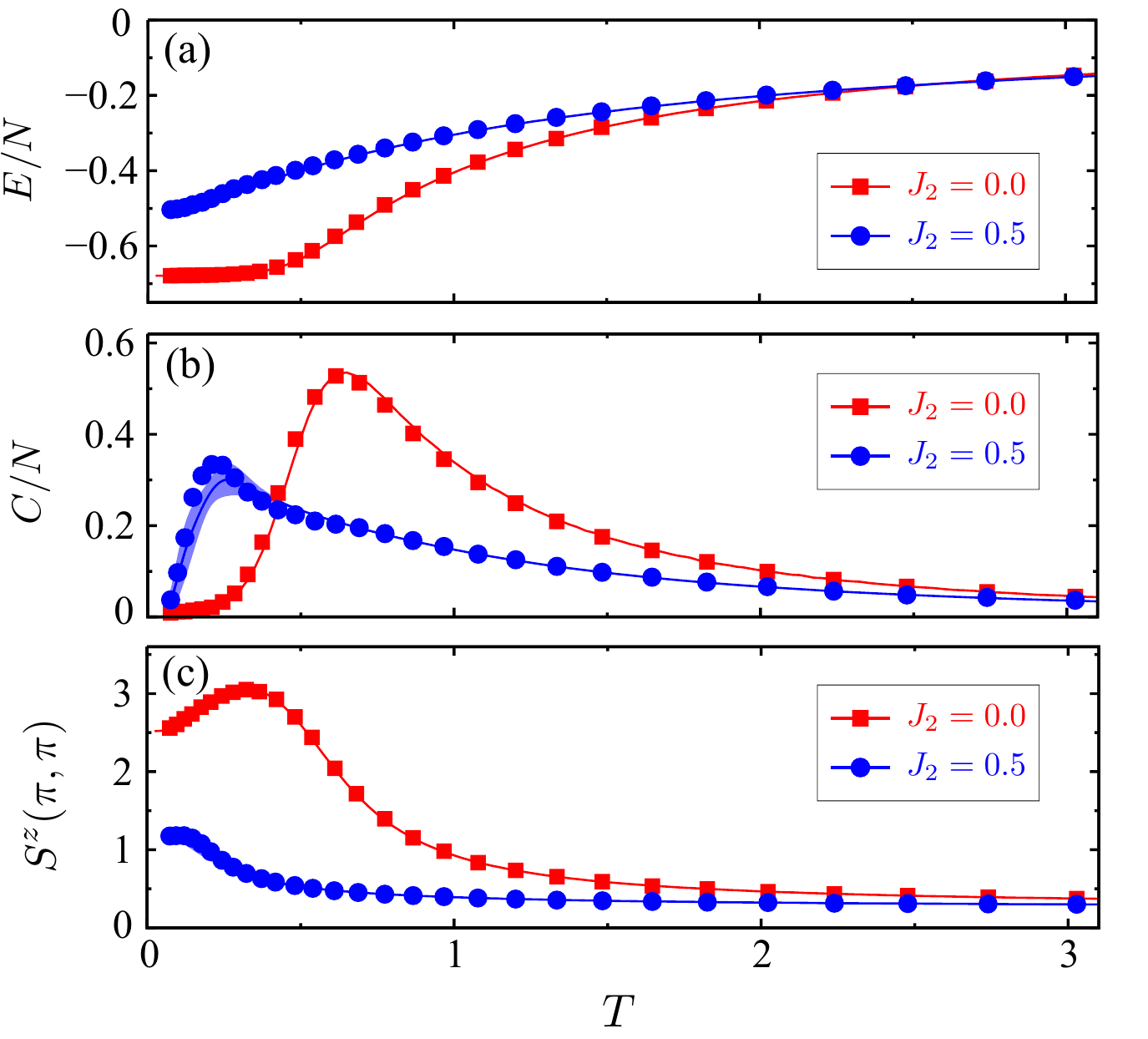}
\caption{
Finite-temperature simulations of the $J_1$--$J_2$ Heisenberg model ($J_1=1$) on the $6\times6$ square lattice with periodic boundary condition. 
(a) Energy, (b) specific heat, and (c) spin structure factor at momentum $(\pi,\pi)$. 
Symbols indicate results obtained by Method (ii) with $M/N=8$, which show good agreement with numerically exact references (solid curves). Shaded regions show the size of the error bars of the exact references. 
Adapted with permission from Ref.~\cite{Nomura_2021_PRL}. Copyright 2021 by the American Physical Society.
}
\label{Fig_FiniteT_J1J2}
\end{center}
\end{figure}

Methods (i) and (ii) have a relationship similar to that of the DBM and RBM algorithms for zero-temperature calculations (Table~\ref{table:RBM_DBM}).
While Method (i) has the advantage of not requiring cumbersome numerical optimization of DBM parameters, 
it requires Monte Carlo sampling for $\sigma$, $\sigma'$, $h$, and $d$ spin configurations to calculate physical quantities, which may cause negative sign problems in general quantum systems.
On the other hand, in Method (ii), the analytical form of the wave function of the extended system $\Psi(\sigma,\sigma')$ can be obtained, and the physical quantities can be calculated by Monte Carlo sampling using only the $\sigma$ and $\sigma'$ configurations with the square of the wave function $|\Psi(\sigma,\sigma')|^2$ as the weight.
This allows Method (ii) to be applied to frustrated spin systems.
In the benchmark calculations in Ref.~\cite{Nomura_2021_PRL}, it was shown that numerically exact solutions can be obtained using Method (i) for frustration-free quantum spin models that do not suffer from the negative sign problem, and that accurate finite-temperature calculations can be performed using Method (ii) for frustrated quantum spin models (Fig.~\ref{Fig_FiniteT_J1J2}).

\section{Summary and Outlook}
\label{Sec_Summary}

This paper presents a method for embedding quantum correlations of quantum many-body systems at zero and finite temperatures in Boltzmann machines.
Using DBMs, one can analytically reproduce the imaginary-time Hamiltonian evolution of quantum states, thereby proving the versatile representative power of the DBM. 
This provides a novel method of quantum-to-classical mapping, but Monte Carlo sampling for the calculation of physical quantities suffers from negative sign problems when the method is applied to, e.g., frustrated spin systems. 
When negative signs become severe in this quantum-to-classical mapping method, the variational approach using RBM becomes useful. 
In the variational approach, at the moment, one needs to pay attention to how to construct a variational ansatz. 
With proper setup, the machine learning method can achieve state-of-the-art accuracy. 
This state-of-the-art variational method has begun to contribute to solving challenging quantum many-body problems.

In this way, we start to understand that the machine learning methods (+ appropriate setups) are useful also in quantum physics. 
In addition to the obvious need for further method improvement, a key question for the future is whether unknown and non-trivial physical insights can be drawn from neural network methods, beyond a black box that somehow yields good results. 
It will be important to explore the boundary between the neural network methods and the physical intuition that essential physics is described by a small number of parameters.

Another important direction is cross-checking using various variational algorithms, including neural network methods. 
For example, in the case of the two-dimensional $J_1$--$J_2$ Heisenberg model, several studies using different numerical algorithms show at least qualitative agreement in the ground-state phase diagram~\cite{Wang_2018,Ferrari_2020,Nomura_2021_PRX,Liu_2022}. 
This makes the results more reliable.
Such cross-checking will become increasingly important when the analytical solution is not available. 
To make comparison among different numerical algorithms meaningful, a consistent accuracy metric is desired.
With this in mind, attempts are being made to build a platform for comparing different algorithms~\cite{Wu_arXiv}.
Further development of variational approaches, including both classical and quantum algorithms, and reliable comparisons between methods will certainly contribute to a deeper understanding of quantum many-body physics.

\ack
We acknowledge fruitful discussions with A. S. Darmawan, Y. Yamaji, M. Imada, G. Carleo, N. Yoshioka, and F. Nori.  
This work was supported by Grant-in-Aids for Scientific Research (JSPS KAKENHI) [Grant Numbers JP23H04869, JP23H04519, JP23K03307, and JP21H01041], MEXT as ``Program for Promoting Researches on the Supercomputer Fugaku'' (Grant Number JPMXP1020230411), and JST (Grant Number JPMJPF2221).

\appendix
\section{Details of SR method}

\subsection{Derivation of parameter update rule}
\label{sec_SR_derivation}

Here, we derive the equation for the parameter update [Eq.~(\ref{Eq:SR_update})] from Eq.~(\ref{Eq_SR_goal}).
Eq.~(\ref{Eq_SR_goal}) is equivalent to
\begin{eqnarray}
\quad    \delta \widetilde{\theta} = \mathop{\rm arg~max}\limits_{\delta\theta} F^2[e^{-2\delta_\tau \mathcal{H}}\ket{\psi_{\theta}}, \ket{\psi_{\theta + \delta \theta}}].
    \label{Eq:F2_SR}
\end{eqnarray}
 Here, $F^2 [e^{-2\delta_\tau \mathcal{H}}\ket{\psi_{\theta}}, \ket{\psi_{\theta + \delta \theta}}]$ is the fidelity between $e^{-2\delta_\tau \mathcal{H}}\ket{\psi_{\theta}}$ and $\ket{\psi_{\theta + \delta \theta}}$:    
\begin{eqnarray}
  \quad  F^2[e^{-2\delta_\tau \mathcal{H}}\ket{\psi_{\theta}}, \ket{\psi_{\theta + \delta \theta}}]  \nonumber \\
   \quad      \quad     =  \frac{ \langle \psi_\theta | e^{-2\delta_\tau \mathcal{H}} |  \psi_{\theta+\delta \theta} \rangle \langle\psi_{\theta+\delta \theta} | e^{-2\delta_\tau \mathcal{H}} | \psi_\theta \rangle }
    {\braket{ \psi_\theta | e^{-4\delta_\tau \mathcal{H}} |  \psi_\theta }  \braket{ \psi_{\theta+\delta \theta} |  \psi_{\theta+\delta \theta} } }.
 \end{eqnarray}
Expanding this equation to the second order for minute quantities yields
\begin{eqnarray}
 \quad   F^2 = 1 - \biggl (&&\sum_{k, l}\delta \theta_k  S_{kl} \delta \theta_l  +2 \delta_\tau \sum_k  g_k \delta \theta_k   + 4 \delta_\tau^2 E_{\rm var}    \biggr).  \nonumber \\
 \label{Eq:sqfid}
\end{eqnarray}
Here, $S_{kl}$ is the Fubini-Study metric tensor [see Eqs.~(\ref{Eq:smat_def}) and (\ref{Eq:qmat_def})]. $g_k$ is the gradient of energy with respect to the variational parameter [Eq.~(\ref{Eq:gradient})], whose detailed form reads 
\begin{eqnarray} 
  \quad  g_k =    
    2 {\rm Re} \left( \frac{\bra{\psi_\theta} \mathcal{H} \ket{\partial_k  \psi_\theta}}{\braket{\psi_\theta | \psi_\theta}}  -   \frac{\bra{\psi_\theta} \mathcal{H} \ket{\psi_\theta}}{\braket{\psi_\theta | \psi_\theta}}  \frac{ \braket{ \psi_\theta | \partial_k\psi_\theta} }{\braket{\psi_\theta | \psi_\theta}}   \right ).
       \nonumber  \\ 
\end{eqnarray}
$E_{\rm var}$ is the variance of the total energy 
\begin{eqnarray}
\quad E_{\rm var}  = 
\frac{\bra{\psi_\theta} \mathcal{H}^2 \ket{\psi_\theta}}{\braket{\psi_\theta | \psi_\theta}}  \! - \! 
 \biggl( \frac{\bra{\psi_\theta} \mathcal{H} \ket{\psi_\theta}}{\braket{\psi_\theta | \psi_\theta}} \biggr)^2
 \!  = 
\langle \mathcal{H}^2\rangle  \! -\!  \langle \mathcal{H} \rangle^2. \nonumber \\ 
\end{eqnarray}
By considering the stationary condition in Eq. (\ref{Eq:sqfid}), we obtain the following equation:    
\begin{eqnarray}
\quad    \delta \theta_k =  - \delta_\tau \sum_l  S^{-1}_{kl}  g_l.
\end{eqnarray}
This equation defines the parameter update in the SR scheme.

\subsection{Monte Carlo sampling}
\label{sec_SR_MC}

The numerical cost of exactly computing physical quantities in the SR scheme grows exponentially with respect to the number of degrees of freedom. 
In practical calculations, we estimate the expectation values using the Monte Carlo method 
as in the variational Monte Carlo (VMC) method~\cite{McMillan_1965,Ceperley_1977,Yokoyama_1987_1,Yokoyama_1987_2}.

We can expand a variational state $| \psi_\theta \rangle$ with a certain basis $\{| x \rangle \}$ as 
\begin{eqnarray}
 \quad  | \psi_\theta \rangle = \sum_x  | x \rangle  \psi_\theta(x). 
\end{eqnarray}
Within this basis, the energy expectation value $\langle \mathcal{H} \rangle$ is given by 
\begin{eqnarray}
\quad \frac{\bra{\psi_\theta} \mathcal{H} \ket{\psi_\theta}}{\braket{\psi_\theta | \psi_\theta}}
&=& \frac{ \sum_{ x x'}  \psi_\theta^* (x)  \mathcal{H}_{ x x'}  \psi_\theta (x') } { \sum_x \left | \psi_\theta (x ) \right |^2 } \\
&=& \frac{ \sum_{x}  \left | \psi_\theta (x) \right |^2 E_{\rm loc}(x) } { \sum_x \left | \psi_\theta (x) \right |^2 }
\end{eqnarray}
with 
$\mathcal{H}_{x x'}= \bra{x}\mathcal{H}\ket{x'} $ and $E_{\rm loc}(x) =  \sum_{x'}  \mathcal{H}_{x x'}  \frac{\psi_\theta (x')}{\psi_\theta(x)} $. 
Thus, we can estimate $\langle \mathcal{H} \rangle$ by the Monte Carlo average of $E_{\rm loc}(x)$, where Monte Carlo samples are generated using the weight proportional to $| \psi_\theta (x)  |^2$.

By defining the diagonal operator $O_{ k } = \sum_\sigma \ket{\sigma} O_k^{\rm loc} (\sigma ) \bra{\sigma}$ with $O_k^{\rm loc} (\sigma ) = \frac{ \partial_k \psi_\theta(\sigma) }{\psi_\theta(\sigma)}$,
we can rewrite the equations for $S_{kl}$ and $g_k$ as 
\begin{eqnarray}
  \quad  S_{kl}    =  {\rm Re} \left( \langle O^{\dagger}_k O_l\rangle - \langle O_k^{\dagger}\rangle \langle O_l \rangle \right)  ,
  \label{Eq.S_appendix}
\end{eqnarray}
and 
\begin{eqnarray}
 \quad   g_k   = 2 {\rm Re} \langle  \mathcal{H} O_k \rangle - 2  \langle \mathcal{H} \rangle  {\rm Re }\langle O_k \rangle, 
 \label{Eq.g_appendix}
\end{eqnarray}
respectively. 
Similarly with the energy expectation value, 
all the expectation values in Eqs. (\ref{Eq.S_appendix}) and (\ref{Eq.g_appendix}) can be estimated using the Monte Carlo method with the weight proportional to $| \psi_\theta (x)  |^2$.
For example,  $\langle O_k \rangle$ given by  
\begin{eqnarray} 
\quad  \langle O_k \rangle = \frac{\braket{\psi_\theta | \partial_k  \psi_\theta}}{\braket{\psi_\theta | \psi_\theta}} = \frac{ \sum_{x}  \left | \psi_\theta (x) \right |^2 O_{\rm loc}(x) } { \sum_x \left | \psi_\theta (x) \right |^2 }
\end{eqnarray}
is computed by the Monte Carlo average of $O_{\rm loc}(x)$.

\section*{References}

\bibliographystyle{iopart-num}
\bibliography{main}

\providecommand{\newblock}{}
\begin{thebibliography}{10}
\expandafter\ifx\csname url\endcsname\relax
  \def\url#1{{\tt #1}}\fi
\expandafter\ifx\csname urlprefix\endcsname\relax\def\urlprefix{URL }\fi
\providecommand{\eprint}[2][]{\url{#2}}

\bibitem{becca_sorella_2017}
Becca F and Sorella S 2017 {\em Quantum Monte Carlo Approaches for Correlated
  Systems\/} (Cambridge University Press)

\bibitem{Bardeen_1957}
Bardeen J, Cooper L~N and Schrieffer J~R 1957 {\em Phys. Rev.\/} {\bf 108}(5)
  1175--1204

\bibitem{Anderson_1973}
Anderson P 1973 {\em Materials Research Bulletin\/} {\bf 8} 153 -- 160 ISSN
  0025-5408

\bibitem{Laughlin_1983}
Laughlin R~B 1983 {\em Phys. Rev. Lett.\/} {\bf 50}(18) 1395--1398

\bibitem{Clark_2018}
Clark S~R 2018 {\em Journal of Physics A: Mathematical and Theoretical\/} {\bf
  51} 135301

\bibitem{Huang_2021}
Huang Y and Moore J~E 2021 {\em Phys. Rev. Lett.\/} {\bf 127}(17) 170601

\bibitem{Carleo_2017}
Carleo G and Troyer M 2017 {\em Science\/} {\bf 355} 602--606 ISSN 0036-8075

\bibitem{Smolensky_1986}
Smolensky P 1986 {\em Parallel Distributed Processing: Explorations in the
  Microstructure of Cognition: Foundations\/} (Cambridge: MIT Press)

\bibitem{Cai_2018}
Cai Z and Liu J 2018 {\em Phys. Rev. B\/} {\bf 97}(3) 035116

\bibitem{Liang_2018}
Liang X, Liu W~Y, Lin P~Z, Guo G~C, Zhang Y~S and He L 2018 {\em Phys. Rev.
  B\/} {\bf 98}(10) 104426

\bibitem{Choo_2019}
Choo K, Neupert T and Carleo G 2019 {\em Phys. Rev. B\/} {\bf 100}(12) 125124

\bibitem{Ferrari_2019}
Ferrari F, Becca F and Carrasquilla J 2019 {\em Phys. Rev. B\/} {\bf 100}(12)
  125131

\bibitem{Westerhout_2020}
Westerhout T, Astrakhantsev N, Tikhonov K~S, Katsnelson M~I and Bagrov A~A 2020
  {\em Nat. Commun.\/} {\bf 11} 1593

\bibitem{Szabo_2020}
Szab\'o A and Castelnovo C 2020 {\em Phys. Rev. Research\/} {\bf 2}(3) 033075

\bibitem{Nomura_2021_JPCM}
Nomura Y 2021 {\em J. Phys.: Condens. Matter\/} {\bf 33} 174003

\bibitem{Nomura_2021_PRX}
Nomura Y and Imada M 2021 {\em Phys. Rev. X\/} {\bf 11}(3) 031034

\bibitem{Astrakhantsev_2021}
Astrakhantsev N, Westerhout T, Tiwari A, Choo K, Chen A, Fischer M~H, Carleo G
  and Neupert T 2021 {\em Phys. Rev. X\/} {\bf 11}(4) 041021

\bibitem{M_Li_2022}
Li M, Chen J, Xiao Q, Wang F, Jiang Q, Zhao X, Lin R, An H, Liang X and He L
  2022 {\em IEEE Transactions on Parallel and Distributed Systems\/} {\bf 33}
  2846--2859

\bibitem{Rath_2022}
Rath Y and Booth G~H 2022 {\em Phys. Rev. Res.\/} {\bf 4}(2) 023126

\bibitem{Roth_arXiv}
Roth C, Szabó A and MacDonald A 2023 High-accuracy variational monte carlo for
  frustrated magnets with deep neural networks arXiv:2211.07749

\bibitem{Reh_2023}
Reh M, Schmitt M and G\"arttner M 2023 {\em Phys. Rev. B\/} {\bf 107}(19)
  195115

\bibitem{Chen_arXiv}
Chen A and Heyl M 2023 Efficient optimization of deep neural quantum states
  toward machine precision arXiv:2302.01941

\bibitem{Saito_2017}
Saito H 2017 {\em J. Phys. Soc. Jpn.\/} {\bf 86} 093001

\bibitem{Saito_2018}
Saito H and Kato M 2018 {\em J. Phys. Soc. Jpn.\/} {\bf 87} 014001

\bibitem{Nomura_2017}
Nomura Y, Darmawan A~S, Yamaji Y and Imada M 2017 {\em Phys. Rev. B\/} {\bf
  96}(20) 205152

\bibitem{Luo_2019}
Luo D and Clark B~K 2019 {\em Phys. Rev. Lett.\/} {\bf 122}(22) 226401

\bibitem{Han_2019}
Han J, Zhang L and ${\rm Weinan \ E}$ 2019 {\em Journal of Computational
  Physics\/} {\bf 399} 108929

\bibitem{Choo_2020}
Choo K, Mezzacapo A and Carleo G 2020 {\em Nat. Commun.\/} {\bf 11} 2368

\bibitem{Pfau_2020}
Pfau D, Spencer J~S, Matthews A~G~D~G and Foulkes W~M~C 2020 {\em Phys. Rev.
  Research\/} {\bf 2}(3) 033429

\bibitem{Hermann_2020}
Hermann J, Sch{\"a}tzle Z and No{\'e} F 2020 {\em Nat. Chem.\/} {\bf 12}
  891--897

\bibitem{Stokes_2020}
Stokes J, Moreno J~R, Pnevmatikakis E~A and Carleo G 2020 {\em Phys. Rev. B\/}
  {\bf 102}(20) 205122

\bibitem{Yoshioka_2021}
Yoshioka N, Mizukami W and Nori F 2021 {\em Commun. Phys.\/} {\bf 4} 106

\bibitem{Inui_2021}
Inui K, Kato Y and Motome Y 2021 {\em Phys. Rev. Research\/} {\bf 3}(4) 043126

\bibitem{Moreno_2022}
Moreno J~R, Carleo G, Georges A and Stokes J 2022 {\em Proc. Natl. Acad. Sci.
  USA\/} {\bf 119} e2122059119

\bibitem{Cassella_2023}
Cassella G, Sutterud H, Azadi S, Drummond N~D, Pfau D, Spencer J~S and Foulkes
  W~M~C 2023 {\em Phys. Rev. Lett.\/} {\bf 130}(3) 036401

\bibitem{Nomura_2020}
Nomura Y 2020 {\em J. Phys. Soc. Jpn.\/} {\bf 89} 054706

\bibitem{Deng_2017}
Deng D~L, Li X and Das~Sarma S 2017 {\em Phys. Rev. X\/} {\bf 7}(2) 021021

\bibitem{Deng_2017_2}
Deng D~L, Li X and Das~Sarma S 2017 {\em Phys. Rev. B\/} {\bf 96}(19) 195145

\bibitem{Glasser_2018}
Glasser I, Pancotti N, August M, Rodriguez I~D and Cirac J~I 2018 {\em Phys.
  Rev. X\/} {\bf 8}(1) 011006

\bibitem{Sirui_2019}
Lu S, Gao X and Duan L~M 2019 {\em Phys. Rev. B\/} {\bf 99}(15) 155136

\bibitem{Kaubruegger_2018}
Kaubruegger R, Pastori L and Budich J~C 2018 {\em Phys. Rev. B\/} {\bf 97}(19)
  195136

\bibitem{Choo_2018}
Choo K, Carleo G, Regnault N and Neupert T 2018 {\em Phys. Rev. Lett.\/} {\bf
  121}(16) 167204

\bibitem{Hendry_2019}
Hendry D and Feiguin A~E 2019 {\em Phys. Rev. B\/} {\bf 100}(24) 245123

\bibitem{Vieijra_2020}
Vieijra T, Casert C, Nys J, De~Neve W, Haegeman J, Ryckebusch J and Verstraete
  F 2020 {\em Phys. Rev. Lett.\/} {\bf 124}(9) 097201

\bibitem{Czischek_2018}
Czischek S, G\"arttner M and Gasenzer T 2018 {\em Phys. Rev. B\/} {\bf 98}(2)
  024311

\bibitem{Schmitt_2020}
Schmitt M and Heyl M 2020 {\em Phys. Rev. Lett.\/} {\bf 125}(10) 100503

\bibitem{Nagy_2019}
Nagy A and Savona V 2019 {\em Phys. Rev. Lett.\/} {\bf 122}(25) 250501

\bibitem{Hartmann_2019}
Hartmann M~J and Carleo G 2019 {\em Phys. Rev. Lett.\/} {\bf 122}(25) 250502

\bibitem{Vincentini_2019}
Vicentini F, Biella A, Regnault N and Ciuti C 2019 {\em Phys. Rev. Lett.\/}
  {\bf 122}(25) 250503

\bibitem{Yoshioka_2019}
Yoshioka N and Hamazaki R 2019 {\em Phys. Rev. B\/} {\bf 99}(21) 214306

\bibitem{Irikura_2020}
Irikura N and Saito H 2020 {\em Phys. Rev. Research\/} {\bf 2}(1) 013284

\bibitem{Nomura_2021_PRL}
Nomura Y, Yoshioka N and Nori F 2021 {\em Phys. Rev. Lett.\/} {\bf 127}(6)
  060601

\bibitem{Le_Roux_2008}
Roux N~L and Bengio Y 2008 {\em Neural Comput.\/} {\bf 20} 1631--1649

\bibitem{Montufar_2011}
Montufar G and Ay N 2011 {\em Neural Comput.\/} {\bf 23} 1306--1319

\bibitem{Viteritti_2022}
Viteritti L~L, Ferrari F and Becca F 2022 {\em SciPost Phys.\/} {\bf 12} 166

\bibitem{Jordan_1928}
Jordan P and Wigner E 1928 {\em Z. Physik\/} {\bf 47} 631--651

\bibitem{Bravyi_2002}
Bravyi S~B and Kitaev A~Y 2002 {\em Annals of Physics\/} {\bf 298} 210--226

\bibitem{Gao_2017}
Gao X and Duan L~M 2017 {\em Nat. Commun.\/} {\bf 8} 662

\bibitem{Levine_2019}
Levine Y, Sharir O, Cohen N and Shashua A 2019 {\em Phys. Rev. Lett.\/} {\bf
  122}(6) 065301

\bibitem{Verstraete_2008}
Verstraete F, Murg V and Cirac J 2008 {\em Advances in Physics\/} {\bf 57}
  143--224

\bibitem{Orus_2014}
Or{\'u}s R 2014 {\em Annals of Physics\/} {\bf 349} 117 -- 158

\bibitem{White_1992}
White S~R 1992 {\em Phys. Rev. Lett.\/} {\bf 69}(19) 2863--2866

\bibitem{White_1993}
White S~R 1993 {\em Phys. Rev. B\/} {\bf 48}(14) 10345--10356

\bibitem{Chen_2018}
Chen J, Cheng S, Xie H, Wang L and Xiang T 2018 {\em Phys. Rev. B\/} {\bf
  97}(8) 085104

\bibitem{Affleck_1987}
Affleck I, Kennedy T, Lieb E~H and Tasaki H 1987 {\em Phys. Rev. Lett.\/} {\bf
  59}(7) 799--802

\bibitem{Lu_2019}
Lu S, Gao X and Duan L~M 2019 {\em Phys. Rev. B\/} {\bf 99} ISSN 2469-9969

\bibitem{Pei_2021}
Pei M~Y and Clark S~R 2021 {\em Entropy\/} {\bf 23} ISSN 1099-4300

\bibitem{Melko_2019}
Melko R~G, Carleo G, Carrasquilla J and Cirac J~I 2019 {\em Nat. Phys.\/} {\bf
  15} 887--892

\bibitem{Sorella_2001}
Sorella S 2001 {\em Phys. Rev. B\/} {\bf 64}(2) 024512

\bibitem{Kingma_2014}
Kingma D~P and Ba J 2014 Adam: A method for stochastic optimization
  arXiv:2009.01821

\bibitem{Amari_1992}
Amari S~I, Kurata K and Nagaoka H 1992 {\em IEEE Transactions on Neural
  Networks\/} {\bf 3} 260--271

\bibitem{Amari_1998}
Amari S~I 1998 {\em Neural Comput.\/} {\bf 10} 251--276

\bibitem{Stokes_2020_quantum}
Stokes J, Izaac J, Killoran N and Carleo G 2020 {\em {Quantum}\/} {\bf 4} 269
  ISSN 2521-327X

\bibitem{Rumelhart_1986}
Rumelhart D~E, Hinton G~E and Williams R~J 1986 {\em Nature\/} {\bf 323}
  533--536

\bibitem{Mezzacapo_2009}
Mezzacapo F, Schuch N, Boninsegni M and Cirac J~I 2009 {\em New Journal of
  Physics\/} {\bf 11} 083026

\bibitem{Lubasch_2014}
Lubasch M, Cirac J~I and Ba\~nuls M~C 2014 {\em Phys. Rev. B\/} {\bf 90}(6)
  064425

\bibitem{Gendiar_2002}
Gendiar A and Nishino T 2002 {\em Phys. Rev. E\/} {\bf 65}(4) 046702

\bibitem{Verstraete_2004_arXiv}
Verstraete F and Cirac J~I 2004 Renormalization algorithms for quantum-many
  body systems in two and higher dimensions (\textit{Preprint}
  \eprint{cond-mat/0407066})

\bibitem{Tahara_2008}
Tahara D and Imada M 2008 {\em J. Phys. Soc. Jpn.\/} {\bf 77} 114701

\bibitem{Gutzwiller_1963}
Gutzwiller M~C 1963 {\em Phys. Rev. Lett.\/} {\bf 10}(5) 159--162

\bibitem{Jastrow_1955}
Jastrow R 1955 {\em Phys. Rev.\/} {\bf 98}(5) 1479--1484

\bibitem{Balents_2010}
Balents L 2010 {\em Nature\/} {\bf 464} 199

\bibitem{Y_Zhou_2017}
Zhou Y, Kanoda K and Ng T~K 2017 {\em Rev. Mod. Phys.\/} {\bf 89} 025003

\bibitem{Hu_2013}
Hu W~J, Becca F, Parola A and Sorella S 2013 {\em Phys. Rev. B\/} {\bf 88}
  060402(R)

\bibitem{Gong_2014}
Gong S~S, Zhu W, Sheng D~N, Motrunich O~I and Fisher M~P~A 2014 {\em Phys. Rev.
  Lett.\/} {\bf 113}(2) 027201

\bibitem{Misawa_2019}
Misawa T, Morita S, Yoshimi K, Kawamura M, Motoyama Y, Ido K, Ohgoe T, Imada M
  and Kato T 2019 {\em Comput. Phys. Commun.\/} {\bf 235} 447--462

\bibitem{Carleo_2018}
Carleo G, Nomura Y and Imada M 2018 {\em Nat. Commun.\/} {\bf 9} 5322

\bibitem{Freitas_2018}
Freitas N, Morigi G and Dunjko V 2018 {\em Int. J. Quantum Inf.\/} {\bf 16}
  1840008

\bibitem{Wang_2018}
Wang L and Sandvik A~W 2018 {\em Phys. Rev. Lett.\/} {\bf 121} 107202

\bibitem{Ferrari_2020}
Ferrari F and Becca F 2020 {\em Phys. Rev. B\/} {\bf 102}(1) 014417

\bibitem{Liu_2022}
Liu W~Y, Gong S~S, Li Y~B, Poilblanc D, Chen W~Q and Gu Z~C 2022 {\em Sci.
  Bull.\/} {\bf 67} 1034--1041

\bibitem{Wu_arXiv}
Wu D, Rossi R, Vicentini F, Astrakhantsev N, Becca F, Cao X, Carrasquilla J,
  Ferrari F, Georges A, Hibat-Allah M, Imada M, Läuchli A~M, Mazzola G,
  Mezzacapo A, Millis A, Moreno J~R, Neupert T, Nomura Y, Nys J, Parcollet O,
  Pohle R, Romero I, Schmid M, Silvester J~M, Sorella S, Tocchio L~F, Wang L,
  White S~R, Wietek A, Yang Q, Yang Y, Zhang S and Carleo G 2023 Variational
  benchmarks for quantum many-body problems arXiv:2302.04919

\bibitem{McMillan_1965}
McMillan W~L 1965 {\em Phys. Rev.\/} {\bf 138}(2A) A442--A451

\bibitem{Ceperley_1977}
Ceperley D, Chester G~V and Kalos M~H 1977 {\em Phys. Rev. B\/} {\bf 16}(7)
  3081--3099

\bibitem{Yokoyama_1987_1}
Yokoyama H and Shiba H 1987 {\em J. Phys. Soc. Jpn.\/} {\bf 56} 1490--1506

\bibitem{Yokoyama_1987_2}
Yokoyama H and Shiba H 1987 {\em J. Phys. Soc. Jpn.\/} {\bf 56} 3582--3592

\end{thebibliography}

\end{document}